\documentclass[lettersize,journal]{IEEEtran}
\usepackage{amsmath,amsfonts}
\usepackage{algorithmic}
\usepackage{algorithm}
\usepackage{amssymb}
\usepackage{multirow}
\usepackage{array}
\usepackage[caption=false,font=normalsize,labelfont=sf,textfont=sf]{subfig}
\usepackage{textcomp}
\usepackage{stfloats}
\usepackage{url}
\usepackage{verbatim}
\usepackage{graphicx}
\usepackage{cite}
\usepackage{xcolor}

\makeatletter

\newcommand{\Rmnum}[1]{\expandafter\@slowromancap\romannumeral #1@}
\makeatother
\begin{document}

\title{\huge Time-Domain Channel Measurements and Small-Scale \\Fading Characterization for RIS-Assisted Wireless Communication Systems}

\author{Yanqing Ren, Mingyong Zhou, Xiaokun Teng, Shengguo Meng, Wankai Tang,~\IEEEmembership{Member,~IEEE,} \\Xiao Li,~\IEEEmembership{Member,~IEEE,} Shi Jin,~\IEEEmembership{Fellow,~IEEE,} and Michail Matthaiou,~\IEEEmembership{Fellow,~IEEE}

\thanks{This work was supported in part by the National Key Research and Development Program of China 2023YFB3811505, 2018YFA0701602, in part by the National Natural Science Foundation of China (NSFC) under Grants 62261160576, 62231009 and 62201138, in part by the Key Technologies R\&D Program of Jiangsu (Prospective and Key Technologies for Industry) under Grants BE2023022 and BE20230221, in part by the Natural Science Foundation of Jiangsu Province under Grant BK20220809, and in part by the European Research Council (ERC) under the European Union’s Horizon 2020 research and innovation programme (grant agreement No. 101001331).
\textit{(Corresponding authors: Shi Jin and Wankai Tang)}
	
Y. Ren, M. Zhou, X. Teng, S. Meng, W. Tang, X. Li, and S. Jin are with the National Mobile Communications Research Laboratory, Southeast University, Nanjing 210096, China. (e-mail: renyq@seu.edu.cn, myzhou@seu.edu.cn, xkteng@seu.edu.cn, seumengsg@seu.edu.cn, tangwk@seu.edu.cn, li\_xiao@seu.edu.cn, and jinshi@seu.edu.cn).

M. Matthaiou is with the Centre for Wireless Innovation (CWI), Queen’s University Belfast, U.K..
(email: m.matthaiou@qub.ac.uk).}}


\IEEEpubid{\begin{minipage}{\textwidth}\ \centering
		Copyright \copyright 20xx IEEE. Personal use of this material is permitted. \\
		However, permission to use this material for any other purposes must be obtained 
		from the IEEE by sending an email to pubs-permissions@ieee.org.
\end{minipage}}

\maketitle

\begin{abstract}

Reconfigurable intelligent surfaces (RISs) have attracted extensive attention from industry and academia. In RIS-assisted wireless communication systems, practical channel measurements and modeling serve as the foundation for system design, network optimization, and performance evaluation. In this paper, a RIS time-domain channel measurement system, based on a software defined radio platform, is developed for the first time to investigate the small-scale fading characteristics of RIS-assisted channels. We present RIS channel measurements in corridor and laboratory scenarios and compare the power delay profile of the channel without RIS, with RIS specular reflection, and with RIS intelligent reflection. The multipath component parameters and cluster parameters based on the Saleh–Valenzuela model are extracted. We find that the power delay profiles (PDPs) of the RIS-assisted channel fit the power-law decay model better than the common exponential decay model and approximate the law of square decay. Through intelligent reflection, the RIS can decrease the delay and concentrate the energy of the virtual line-of-sight (VLoS) path, thereby reducing the delay spread and mitigating multipath fading. Furthermore, the cluster characteristics of RIS-assisted channels are highly dependent on the measurement environment. In the laboratory scenario, a single cluster dominated by the VLoS path with smooth envelope is observed. On the other hand, in the corridor scenario, some additional clusters introduced by the RIS reflection are created.
\end{abstract}


\section{Introduction}
\IEEEPARstart{I}{n} recent years, fifth generation (5G) communication systems have become widely deployed and commercially available in the world \cite{shafi_5g_2017}. The research, exploration and evaluation of the potential key technologies of the sixth generation (6G) communication systems have already commenced. 6G puts higher expectations on the performance of mobile communication systems, such as ultra-high speed, ultra-high energy efficiency, and air-ground-sea cooperative networking \cite{Matthaiou:COMMag:2021}. On one hand, in order to enhance the network coverage and improve the network capacity, the number of active nodes, such as base stations (BSs) with massive antennas, will increase significantly, which may in principle increase the energy consumption and hardware cost, as well as the signal processing burden \cite{zhang2020prospective}. On the other hand, high-frequency band communications, such as millimeter-wave (mmWave) and terahertz bands, inherently suffer from severe path loss and penetration losses, which entails serious challenges for achieving full-coverage communications \cite{chen_survey_2019}. Therefore, 6G is actively seeking new paradigms for wireless communication systems in response to the aforementioned demands and problems. Among them, reconfigurable intelligent surfaces (RISs) have attracted a lot of attention for their low-cost, low-power, programmable, and easy-to-deploy properties \cite{han_large_2019,9340586,direnzo_smart_2020,wu_intelligent_2021}.

\IEEEpubidadjcol

With the capability of reconfiguring the propagation environment, RISs have found a wide range of practical applications, such as coverage enhancement for indoor and outdoor scenes \cite{sang_coverage_2022,li_coverage_2022}, joint beamforming for improving the achievable rate and reception reliability \cite{feng_deep_2020,feng_joint_2021}, as well as assisting vehicular communication networks \cite{shaikh_performance_2022,ai_secure_2021,huang_transforming_2022,9690475}. In RIS-aided vehicle-to-infrastructure (V2I) communications, RISs can be used to improve the quality and secrecy of vehicular communications by establishing low-cost, highly energy-efficient indirect line-of-sight (LoS) links \cite{shaikh_performance_2022,ai_secure_2021}. Furthermore, on-vehicle RISs can enable the mitigation of small-scale fading of the BS-user channels for high-mobility communications \cite{huang_transforming_2022}. It is envisioned that future wireless networks for supporting RIS-aided vehicular communication will have an integrated 3D architecture consisting of trackside-based RISs, window-based RISs, and wall-based RISs. Specifically, a wall-based RIS can be attached to the wall of vehicles to enhance the received signal, similar to a traditional indoor scenario \cite{9690475}.

The potential suitability of RIS for mmWave/sub-THz propagation is well understood. However, in low frequency bands, such as the sub-6 GHz band, RISs also have great potential for coverage enhancement and capacity boosting of wireless networks, underpinned by their ability to customize wave propagation and to achieve smart radio environments \cite{direnzo_smart_2020}. In the sub-6 GHz band, RISs demonstrated significant coverage enhancement capabilities in current 5G commercial mobile networks \cite{sang_coverage_2022,li_coverage_2022}. In addition, a RIS is able to actively enrich the channel scattering conditions to enhance the multiplexing gain of multiple-input multiple-output (MIMO) wireless communication at 2.6 GHz and 5 GHz \cite{dunna_scattermimo_2020,meng_rank_2023}. In low-frequency scenarios with rich scattering environments, RISs enable the shaping of the multipath channel impulse response on demand and the localization of non-cooperative objects using wave fingerprints \cite{alexandropoulos_reconfigurable_2021}.

The study of RIS channel characteristics is important for the design, simulation and performance evaluation of RIS-assisted wireless communication systems. Various works have been dedicated to exploring the RIS channel large-scale properties through channel measurements and modeling. In \cite{tang_wireless_2021}, free-space path loss models for RIS-assisted wireless communications were developed for various applications of far-field beamforming, near-field beamforming, and near-field broadcasting. Experimental measurements were conducted in a microwave anechoic chamber by using three fabricated RISs to further corroborate the theoretical models. Furthermore, \cite{tang_path_2022} refined the previously free-space pass loss models and validated their accuracy in the sub-6 GHz and millimeter-wave bands by channel measurements. In order to model the RIS path loss in real scenarios, \cite{gao_propagation_2022} modified the floating-intercept (FI) model and the close-in (CI) model. The path loss and the shadow fading of the RIS-assisted communication system were measured and analyzed in a corridor using a vector network analyzer (VNA). Considering the waveguide effect in corridors, \cite{li_path_2022} carried out channel measurements in a typical indoor corridor scenario and proposed a dual-slope path loss model suitable for the RIS-assisted channel in a waveguide-like structure scenario. 

Nevertheless, to the best of our knowledge, studies on small-scale fading characteristics of RIS wireless channels are still in their infancy. In this context, \cite{zhang_channel_2023} proposed that RIS-assisted channels may exhibit characteristics similar to those of massive MIMO, such as near field effects and spatial non-stationarity. Moreover, \cite{sun_3d_2021} developed an extended geometry-based stochastic model (GBSM) for RIS-assisted channels and verified it via a spatial autocorrelation function. However, different from theoretical research, channel measurements and parameter extraction of small-scale fading in real environments are indispensable for small-scale channel modeling, which, however, have been only sporadically reported in the literature. In this context, \cite{sang_multiscenario_2023} measured and analyzed the propagation characteristics in outdoor, indoor, and outdoor-to-indoor (O2I) environments for RIS-assisted channels, including large-scale fading, such as path loss and time-frequency-space characteristics. However, from the perspective of channel power delay profiles (PDPs), what kind of multipath and clustering phenomena will be brought by virtual line-of-sight (VLoS) paths introduced by a RIS, and their impact on small-scale parameters deserves further exploration.

Based on the above issues, this paper will present time domain channel measurements and characterization of RIS-assisted wireless communication. We build a RIS time-domain channel measurement system and carry out a set of measurements of RIS-assisted channels in indoor scenarios. Based on the acquired measurement data, we extract small-scale channel characteristic parameters and study the impact of the RIS on the channel characteristics, in order to acquire a reference for the design, simulation, and performance evaluation of RIS-assisted wireless communication systems. The specific contributions of this paper are as follows:
 
	\begin{itemize}
		\item{RIS channel measurements are carried out at typical locations in indoor corridor and laboratory scenarios, respectively. The PDPs of three channel modes are analyzed and compared to investigate the impact of RIS, namely without RIS, RIS specular reflection and RIS intelligent reflection. We find that RIS intelligent reflection is effective in introducing VLoS paths with good spatial consistency in PDPs. Additionally, the phenomenon of multiple reflections in the RIS-assisted channel is observed in the corridor scenario and is dependent on the location of the receiver.}
		\item{Based on the tap delay line model, the fading pattern of the PDPs is analyzed. Since the RIS concentrates the energy on the VLoS path, the average PDP of the intelligent reflection is found to fit better with the power-law decay model compared to the common exponential decay model.}
		\item{We extract multipath parameters from the measured PDPs, such as the multipath power, mean delay, and delay spread. It is unveiled that a RIS with intelligent reflection enhances and concentrates the energy of the VLoS path, thus mitigating the multipath fading and reducing the delay spread.}
		\item{We further cluster the measured PDPs based on the Saleh–Valenzuela (S-V) model. We discover that the cluster characteristics of RIS-assisted channels are highly dependent on the measurement environment. Moreover, we observe that the multiple reflected paths introduced by the RIS form reflected clusters with the power-law decaying rays, which have an impact on parameters such as the cluster arrival rate, cluster decay factor, and ray decay factor.}
		
	\end{itemize}

The rest of this paper is organized as follows: Section \Rmnum{2} describes the time-domain measurement principle, the measurement system, and the measurement environment. Section \Rmnum{3} presents the PDP models and analyzes the PDP features in RIS-assisted channels. The delay dispersion parameters and other subsidiary parameters are calculated to further decipher the role of a RIS in the wireless channel. Section \Rmnum{4} adopts the S-V model to depict the cluster characteristics of RIS-assisted channels. A heuristic clustering algorithm is also introduced and the clustering results are analyzed. The conclusions are drawn in Section \Rmnum{5}.

\section{RIS Channel Measurement System and Scenario Setup}

\subsection{The Measurement Principle}

\begin{figure*}[t]
	\centering
	\includegraphics[width=7in]{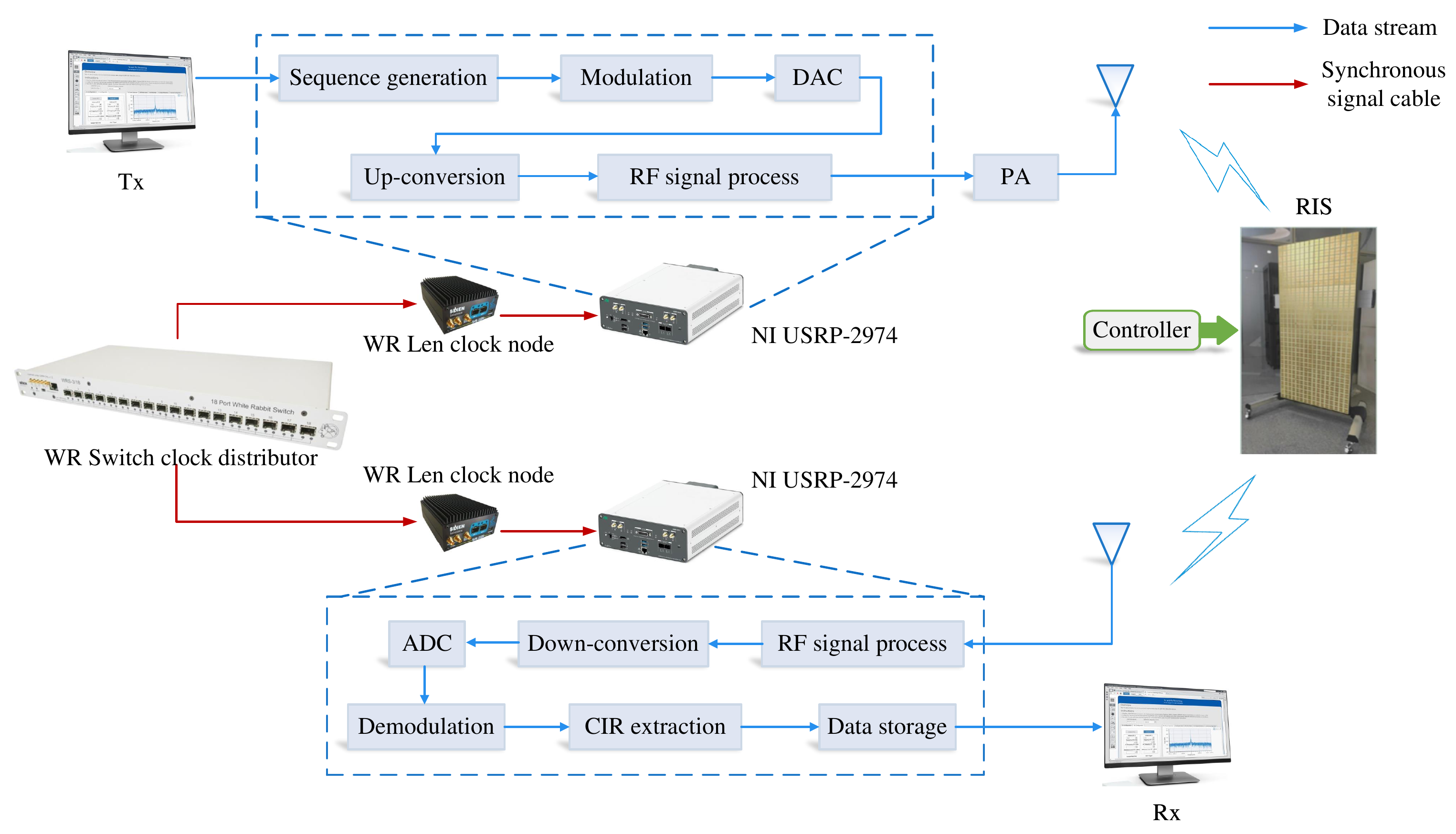}
	\caption{The architecture diagram of the USRP-based RIS time-domain channel measurement system, where the main hardware equipment, the USRP internal data stream and the synchronization module of the system are illustrated.}
	\label{system}
\end{figure*}

We adopt the pseudo-random (PN) sequence correlation method to measure the RIS-assisted channel. Denote the PN sequence as $c_k$, the sequence period as $K_c$, the pulse waveform as $w (t)$, and the pulse width as $T_c$. Then, the PN probe signal $x (t)$ is written as 
\begin{equation}
	\label{xt}
	x\left( t \right) =  \sum\limits_{k = 1}^{K_c} c_k w \left( {t - k{T _c}} \right).
\end{equation}
Propagating through the multipath channel, the baseband received signal can be expressed as
\begin{equation}
	\label{yt}
	y\left( t \right) = x\left( t \right) * h\left( t, \tau \right) + n\left(t\right),
\end{equation}
where $h\left( t, \tau \right)$ is the CIR, $n\left( t \right)$ is additive Gaussian white noise, and $*$ denotes the mathematical operation of convolution. Considering that the channel is static and the transceiver is stationary at each measurement point, the CIR can be formulated as
\begin{equation}
	\label{ht}
	h\left( \tau \right) = \sum\limits_{i = 1}^{N_{\mathrm{path}}} {{A_i}} {e^{j{\phi _i}}}\delta \left( { \tau - {\tau _i}} \right),
\end{equation}
where $N_{\mathrm{path}}$ is the number of multipaths in the wireless channel while the amplitude, phase, and delay of the $i$-th path are denoted as ${A_i}$, $\phi _i$, and $\tau _i$, respectively. At the receiving end, by performing a cross-correlation operation between the received signal $ y\left( t \right)$ and the same PN signal as $ x\left( t \right)$, we can get the estimated CIR
\begin{equation}
	\label{ht^}
	\hat h\left( \tau \right) = \sum\limits_{i = 1}^{N_{\mathrm{path}}} {{A_i}} {e^{j{\phi _i}}}R\left( { \tau - {\tau _i}} \right),
\end{equation}
where  $R(\tau)$ is the autocorrelation function of the PN signal, which is approximated as a delta function $\delta(\tau)$. Thus, $\hat h (\tau) \approx h (\tau)$, after ignoring the noise effect. A back-to-back system calibration is required to exclude the system response.

To further calculate the PDP of the channel, provided that $M$ results are obtained at each measurement point, the PDP is calculated as
\begin{equation}
	\label{pt}
	PDP\left(  \tau \right) = \frac{1}{M}{\sum\limits_{m = 1}^M {\left| {h_m\left( \tau \right)} \right|} ^2},
\end{equation}
where ${h_m\left(\tau \right)}$ is the CIR extracted from the $m$-th measurement. 

\subsection{The Measurement System}

As shown in Fig. \ref{system}, the RIS time-domain channel measurement system is based on the USRP. The transmitter (Tx) site of the system consists of a USRP-2974, a power amplifier, and a transmit antenna. The receiver (Rx) site consists of a USRP-2974 and a receive antenna. The RIS is placed between the Tx and Rx, which can be controlled through a laptop computer. The USRP internally contains baseband and radio frequency (RF) signal processing modules, and is programmed by the LabVIEW software and connected by a monitor to display real-time results. When performing channel measurements, the built-in module of USRP produces PN sequences with length of 511 periodically as the probe sequences. After the modulation of binary phase shift keying (BPSK), the ``0" and ``1" of the PN sequences are mapped to the symbols ``-1" and ``1", respectively. The system sampling rate is set to 200 MHz and the number of sampling points per symbol is 4, resulting in a symbol rate of 50 Baud/s. The modulated symbols are passed through a root-raised cosine filter to produce a continuous pulse waveform. Then, the PN signal experiences digital-to-analog conversion, up-conversion, RF signal processing, power amplification, and is ultimately transmitted from the transmit antenna. After passing through the RIS-assisted channel, the RF signal reaches the receive antenna and undergoes the converse process of RF signal processing, down-conversion, analog-to-digital conversion, demodulation, and CIR extraction. Finally, the CIR data is stored in file form for subsequent data processing and parameter analysis. The main hardware parameters of the system are summarized in Table \ref{table1}.

\begin{table}[t]
	\renewcommand{\arraystretch}{1.5}
	\caption{Main Hardware Parameters of the System\label{table1}}
	\begin{tabular}{|c|c|c|}
		\hline
		\textbf{Equipment} & \textbf{Parameter} & \textbf{Value} \\ \hline
		\multirow{8}{*}{USRP} & Type & USRP-2974 \\ \cline{2-3} 
		& Center frequency & 2.6 GHz \\ \cline{2-3}
		& Probe signal & PN sequence with BPSK \\ \cline{2-3}
		& Signal bandwidth & 75 MHz \\ \cline{2-3}
		& Sampling rate & 200 MHz \\ \cline{2-3} 
		& Delay resolution & 5 ns \\ \cline{2-3}
		& Dynamic range & 40 dB \\ \cline{2-3}		 
		& Transmit power & 0 dBm \\ \hline
		\multirow{5}{*}{RIS} & Operating frequency & 2.6 GHz \\ \cline{2-3} 
		& Size of units & 5 cm × 5 cm \\ \cline{2-3} 
		& Number of units & 32 × 16 \\ \cline{2-3} 
		& Control mode & 1-bit phase programmable \\ \cline{2-3} 
		& Quantification threshold & 55°; 235° \\ \hline
		\multirow{3}{*}{Antenna} & Type & \begin{tabular}[c]{@{}c@{}} Tx: horn antenna \\ Rx: omni-directional \\ antenna \end{tabular} \\ \cline{2-3} 
		& Gain & Tx: 8.25 dBi; Rx: 0 dBi \\ \cline{2-3} 
		& Height & 1.2 m \\ \hline
		Power amplifier & Gain & 30 dB \\ \hline
	\end{tabular}
\end{table}

\begin{figure}[h]
	\centering
	\includegraphics[width=3in]{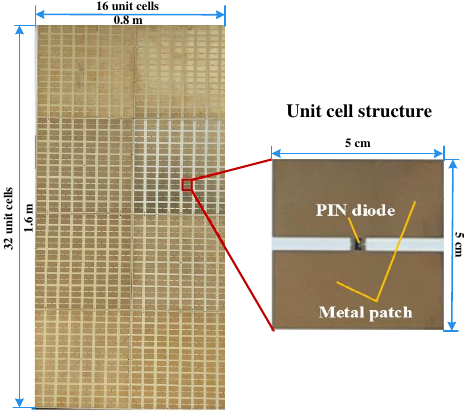}
	\caption{The physical diagram of the used RIS and its unit cell.}
	\label{RIS}
\end{figure}

\begin{figure}[t]
	\centering
	\includegraphics[height=2in]{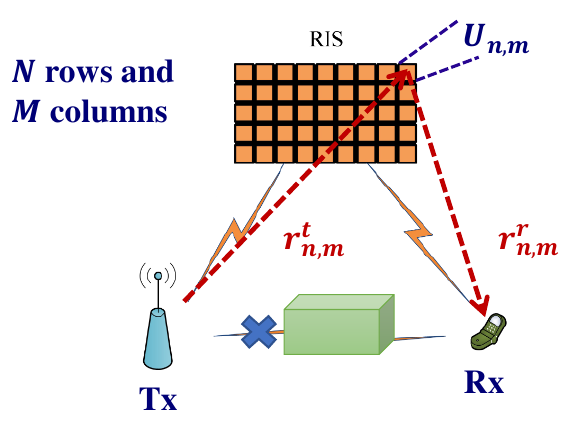}
	\caption{ The schematic diagram of the RIS sub-channel for intelligent reflection.}
	\label{subchannel}
\end{figure}

As illustrated in Fig. \ref{RIS}, the RIS used for our measurements is 1.6 m high, 0.8 m wide, and contains 32 × 16 cells operating at 2.6 GHz. Each unit cell consists of a PIN diode and a pair of metal patches. The cells of the RIS are 1-bit independently phase-programmable and the phase difference between the reflection coefficients of the two states of the cells is 180°. As shown in Fig. \ref{subchannel}, the $(n,m)$-th unit cell of RIS is denoted by ${\rm U}_{n,m}$, where $r_{n,m}^ t$ and $r_{n,m}^ r$ is the distance from the Tx to ${\rm U}_{n,m}$ and the distance from ${\rm U}_{n,m}$ to the Rx, respectively. The optimal continuous phase of $\rm U_{n,m}$ is configured to be [19, Eq. (12)]
	\begin{equation}
		{\phi _{n,m}} = \bmod \left( \frac{ 2\pi \left( r_{n,m}^t + r_{n,m}^r \right) }{\lambda},2\pi \right),
		\label{phi}
	\end{equation}
and subsequently quantized. Specifically, the optimal continuous phase 55° $\sim $ 235° is coded as ``1", and the optimal continuous phase 235° $\sim $ 360° and 0° $\sim$ 55° is coded as ``0". This phase configuration method eliminates the phase differences in sub-channels, actively aligns multipath phases, and concentrates the energy of VLoS path.

A clock distribution system, comprising a White Rabbit (WR) Switch clock distributor and two WR Len clock nodes, is used to achieve synchronization of USRPs at the Tx and Rx. As seen in Fig. \ref{system}, the USRP-2974 at each side is equipped with a WR Len clock node, which is connected to the WR Switch clock distributor via optical fibers. During the channel measurements, the WR Switch clock distributor transmits the 10MHz reference signal to the clock node of the two USRPs through the Ethernet optical port to eliminate the influence of carrier frequency bias and sampling frequency bias.

\subsection{Measurement Scenarios}

\begin{figure}[t]
	\centering
	\subfloat[]{\includegraphics[width=3in]{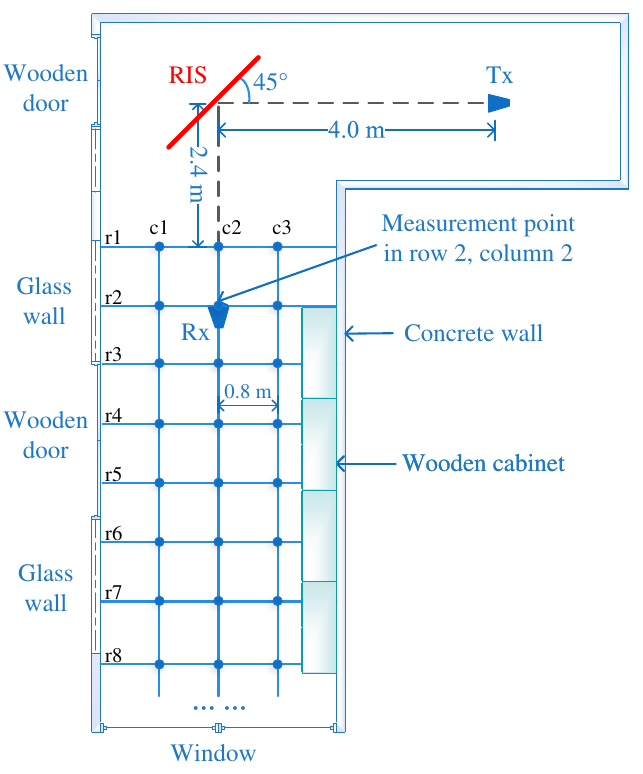}}
	\\
	\subfloat[]{\includegraphics[width=3in]{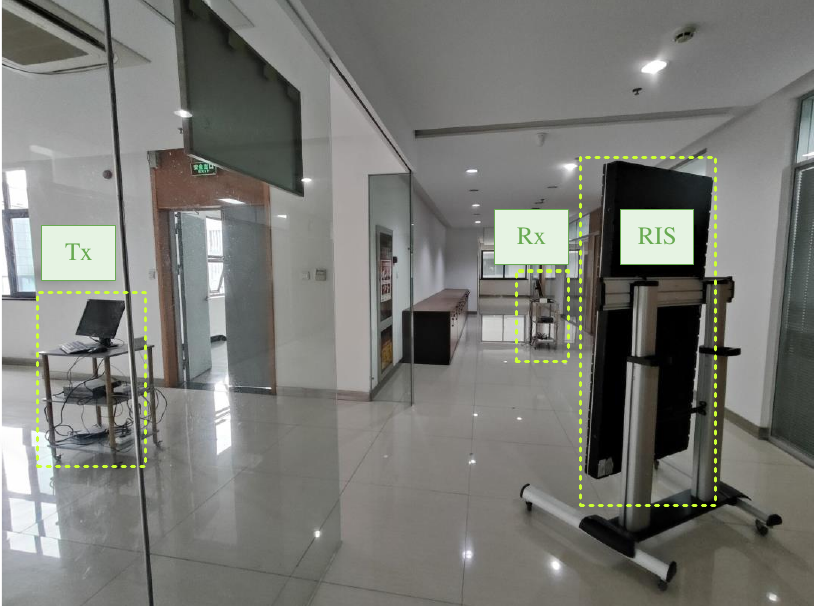}}
	\caption{The channel measurement campaign carried out in the corridor scenario: (a) Schematic diagram of the measurement environment with measurement points marked with blue dots, where ``r$m$" denotes the $m$-th row and ``c$n$" denotes the $n$-th column. (b) Diagram of the real measurement scene.}
	\label{scenario1}
\end{figure}

\begin{figure*}[t]
	\subfloat[]{\includegraphics[width=4.2in]{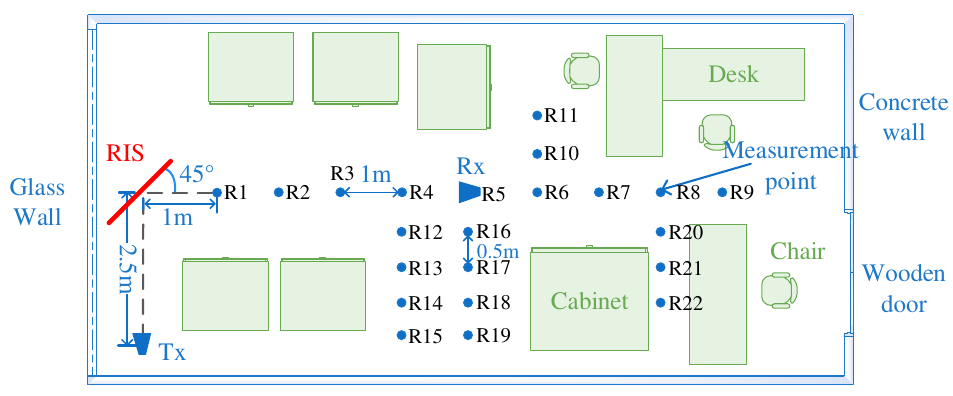}}
	\hfil
	\subfloat[]{\includegraphics[width=2.8in]{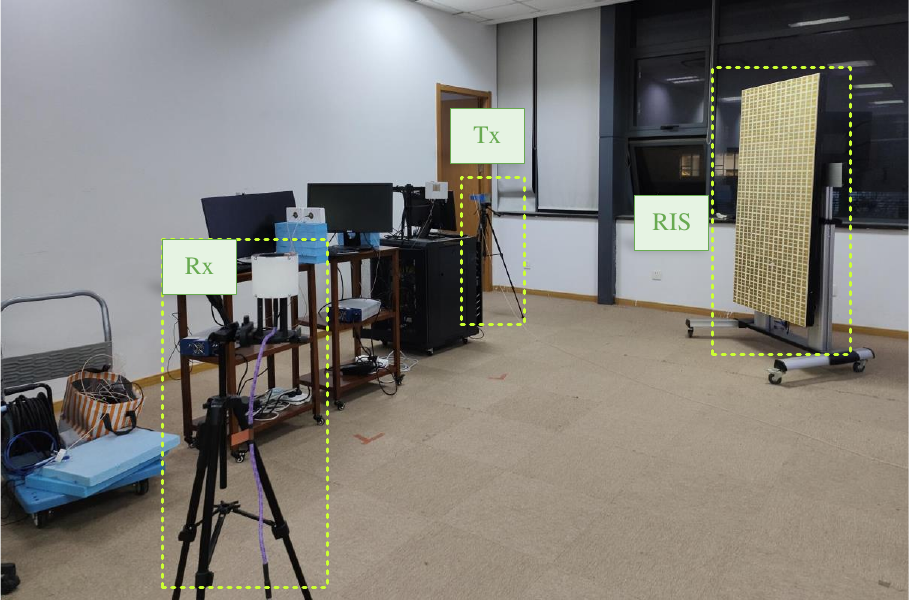}}
	\caption{The channel measurement campaigns carried out in the laboratory scenario: (a) Schematic diagram of the measurement environment with the measurement points marked with ``R(number)". (b) Diagram of the real measurement scene.}
	\label{scenario2}
\end{figure*}

\begin{figure*}[htbp]
	\centering
	\subfloat[]{\includegraphics[width=2.4in]{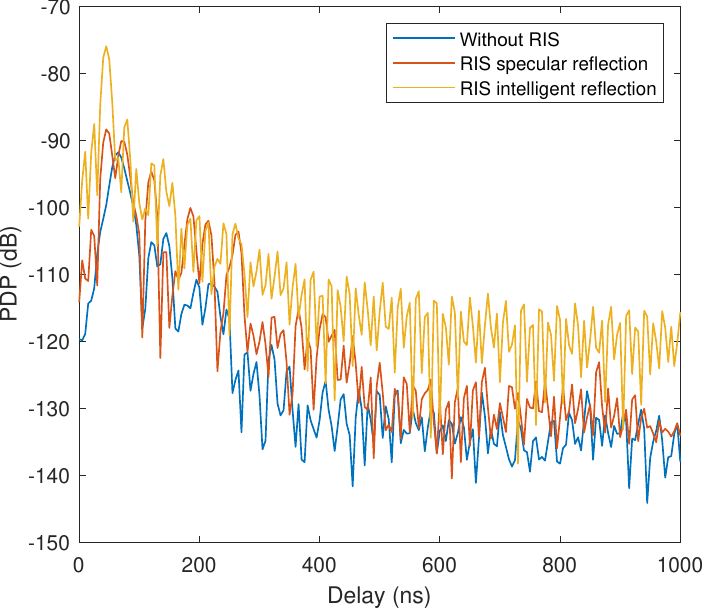}}
	\subfloat[]{\includegraphics[width=2.4in]{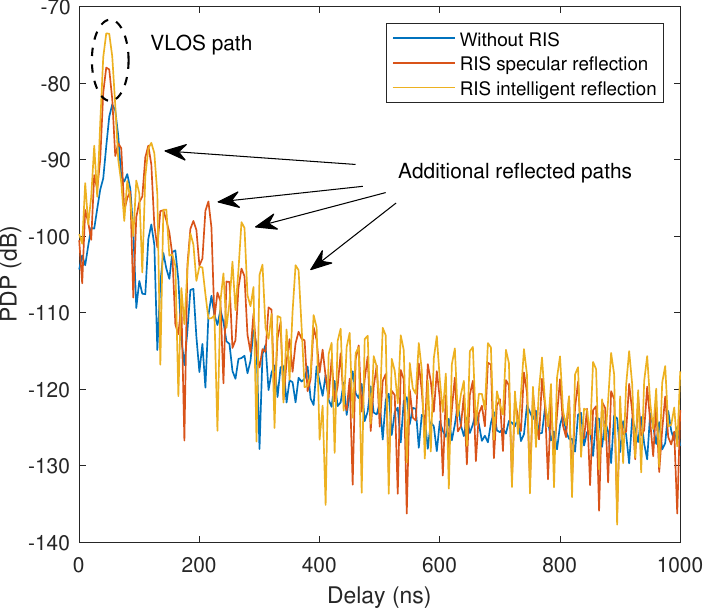}}
	\subfloat[]{\includegraphics[width=2.4in]{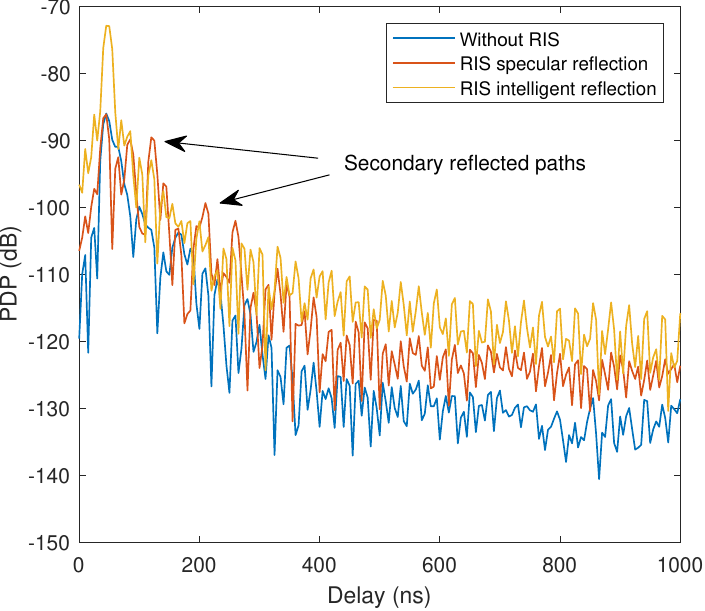}}
	\caption{PDPs of the measurement points at Row 8 in the corridor scenario. The blue curves plot the PDPs in the channel without RIS, the red curves plot the PDPs in the channel with RIS specular reflection, and the orange curves plot the PDPs in the channel with RIS intelligent reflection. (a) PDP at Row 8, Column 1. (b) PDP at Row 8, Column 2. (c) PDP at Row 8, Column 3.}
	\label{pdp1}
	\vspace{-0.3cm}
\end{figure*}

\begin{figure*}[htbp]
	\centering
	\subfloat[]{\includegraphics[width=2.4in]{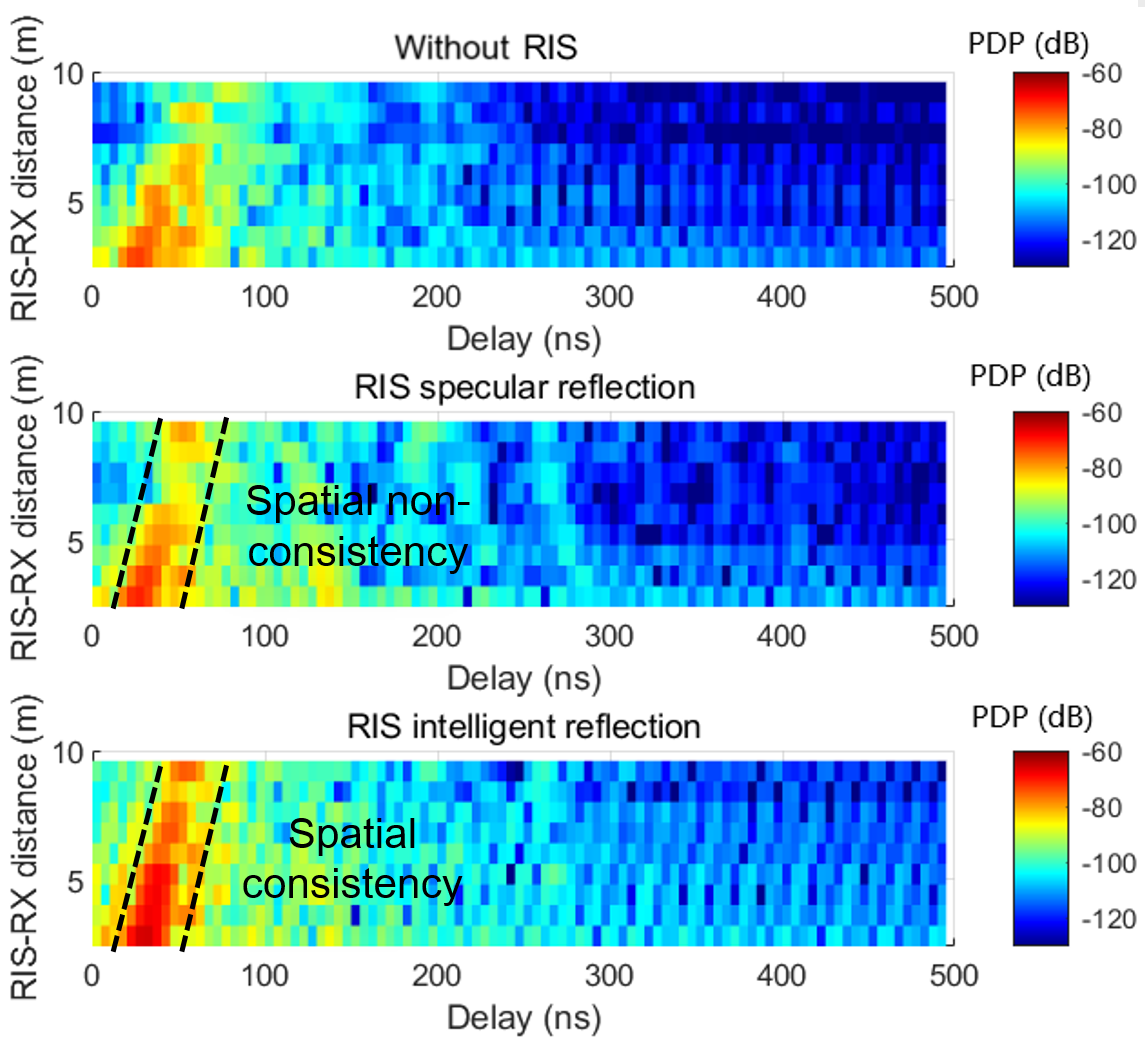}}
	\subfloat[]{\includegraphics[width=2.4in]{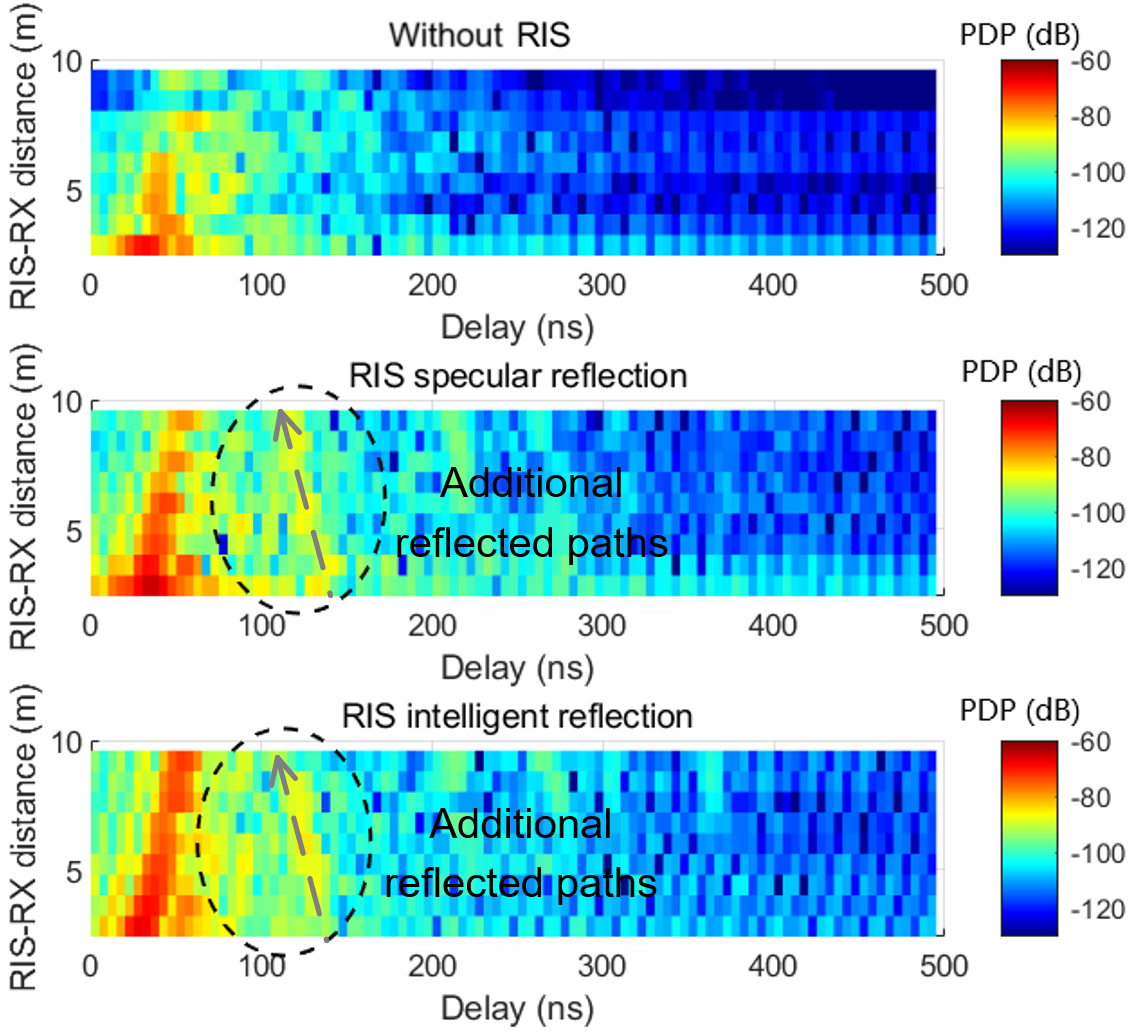}}
	\subfloat[]{\includegraphics[width=2.4in]{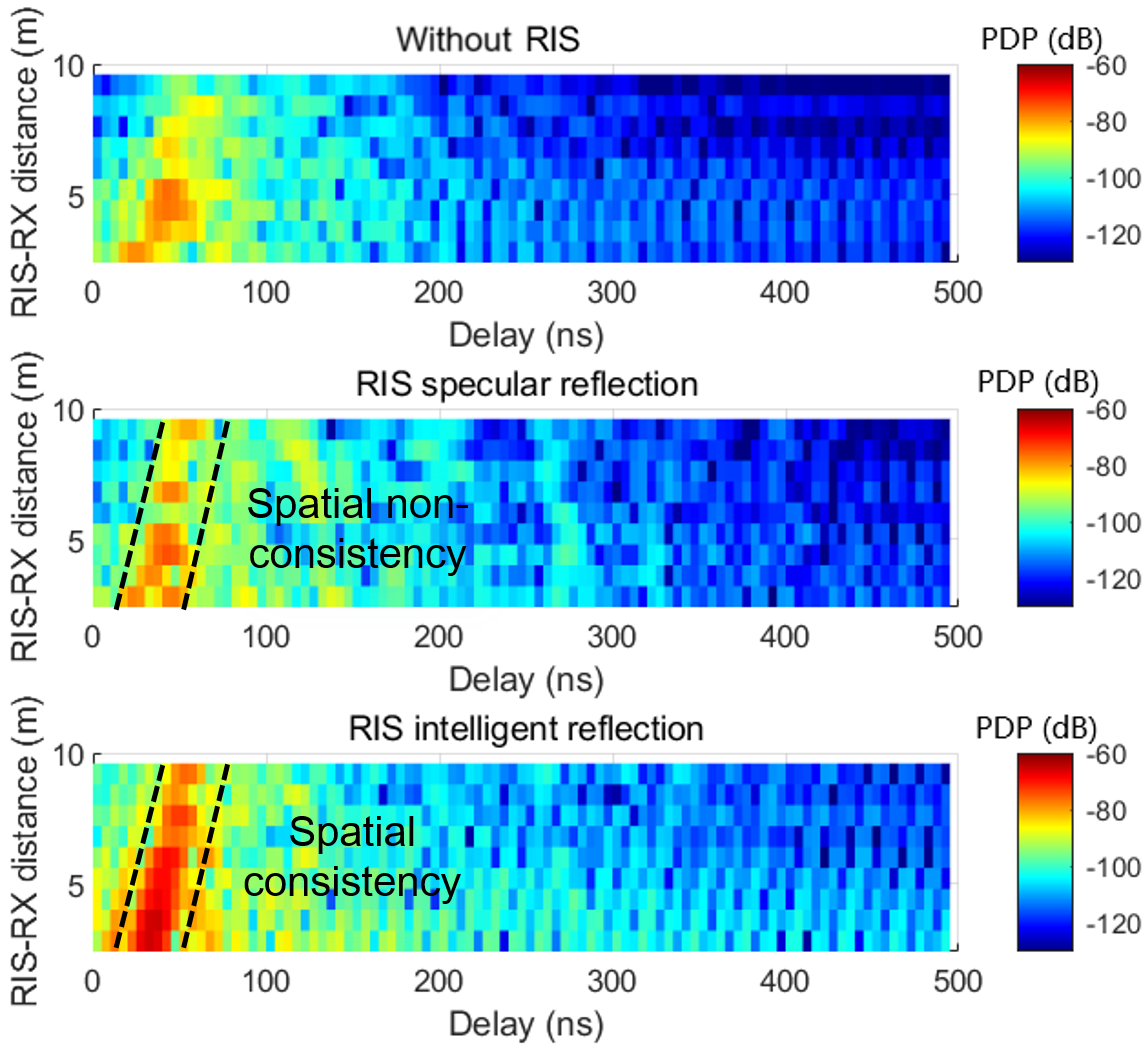}}
	\caption{ Heat maps of the PDP at points in different columns with respect to the RIS-Rx distance in three channel modes in the corridor scenario. (a) PDP evolution versus the RIS-Rx distance in Column 1. (b) PDP evolution versus the RIS-Rx distance in Column 2. (c) PDP evolution versus the RIS-Rx distance in Column~3.}
	\label{heatmap}
\end{figure*}

\begin{figure*}[htbp]
	\centering
	\subfloat[]{\includegraphics[width=2.4in]{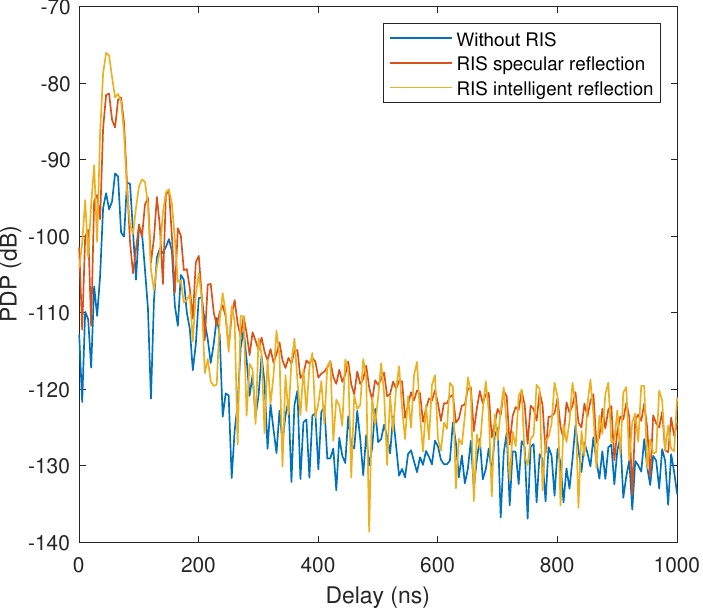}}
	\subfloat[]{\includegraphics[width=2.4in]{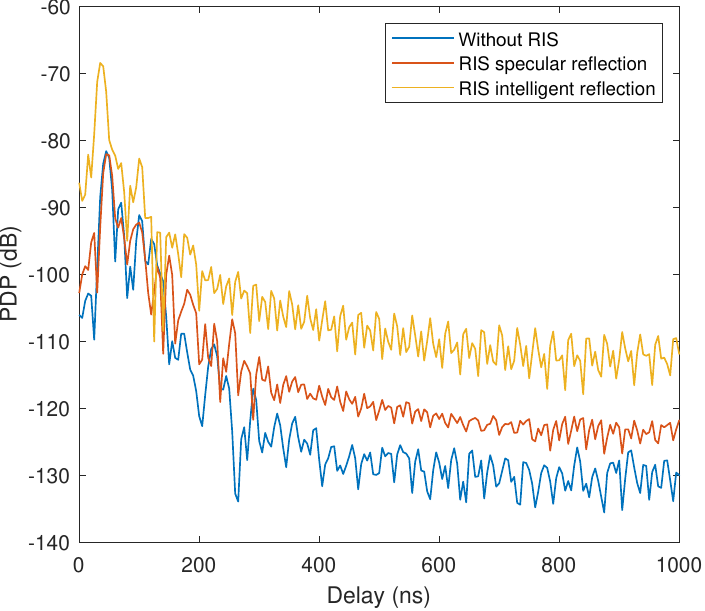}}
	\subfloat[]{\includegraphics[width=2.4in]{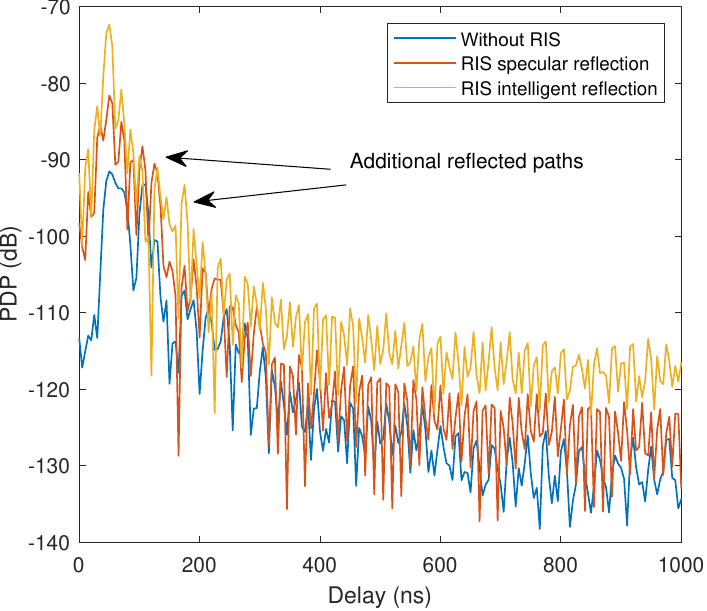}}
	\caption{PDPs of the typical measurement points in the laboratory scenario. The blue curves plot the PDPs in the channel without RIS, the red curves plot the PDPs in the channel with RIS specular reflection, and the orange curves plot the PDPs in the channel with RIS intelligent reflection. (a) PDP at R8. (b) PDP at R19. (c) PDP at R22.}
	\label{pdp2}
	\vspace{-0.3cm}
\end{figure*}

With the help of the RIS time-domain channel measurement system, channel measurement campaigns are carried out in an indoor ``L-shaped" corridor and an indoor laboratory in the third floor of the Nanjing Purple Mountain Laboratories. Three types of channels, \textit{without RIS}, \textit{RIS specular reflection}, and \textit{RIS intelligent reflection} are measured in the two scenarios. Among them, the Tx always transmits the probe signal impinging on the RIS at an angle of 45°, while the Rx moves at different measurement points with distinct reflection directions. In the case of without RIS, there is no RIS placed between the Tx and Rx, which is equivalent to measuring the non-line-of-sight (NLoS) channel at 2.6 GHz. In the case of a RIS specular reflection, the RIS is not coded and acts like a metal plate. The probe signal is reflected by RIS at the same 45° angle. In the case of RIS intelligent reflection, as the Rx moves to each measurement point, we code the RIS so that the RIS beam is focused into the direction of the Rx.

The measurement environment and the arrangement of measurement points in the corridor and the laboratory are shown in Fig. \ref{scenario1} and Fig. \ref{scenario2}, respectively. Both scenarios are NLoS environments. The transmitter deploys a horn antenna and the receiver an omni-directional antenna, which are both at the same height of 1.2 m and oriented to the center of the RIS. During the measurement, the position of the transmit antenna and the RIS are fixed, while the receive antenna is moved in sequence to each measurement point. In order to smooth out the noise, $M=10$ PDPs are recorded at each location of the Rx and the average value is taken as the average PDP of the measurement point.

\subsubsection{The corridor scenario} The long side of the corridor has a total length of 20.8 m and width of 2.9 m. Wooden doors and glass wall are on the left side of the corridor and a row of wooden cabinets about 6.5 m long are placed against the wall on the right side. A glass window with metal frames is at the end of the corridor. The rest of the surfaces are all concrete walls and the floor is covered with marble tiles. There are 10 rows and 3 columns of measurement points and adjacent measurement points are 0.8 m apart from each other, which means the measurement range is from 2.4 m to 9.6 m. The positions of the measurement points are partly marked in Fig. \ref{scenario1}(a) in order to facilitate the analysis of the measurement results in the following, where ``r$m$" denotes the $m$-th row and ``c$n$" denotes the $n$-th column.

\subsubsection{The laboratory scenario} The laboratory is about 12.4 m long and 5.8 m wide. Except for the glass windows on the left side, the other three sides are surrounded by concrete walls, and the floor is covered with the carpet. Desks, chairs, sundries, and the chassis for various experimental equipment are placed against the wall in the laboratory, which together constitute the scattering environment. The transmit antenna is located in the corner of the laboratory, 2.5 m away from the center of RIS. We select 22 measurement points in total, with the spacing of 9 measurement points in the horizontal direction being 1 m, whereas the spacing of the 13 measurement points in the vertical direction is 0.5 m, which are all marked with ``R(number)" in Fig. \ref{scenario2}(a).

\section{PDP and Multipath Parameters in RIS-assisted Channels}

\subsection{PDP Model}

A traditional model for small-scale characterization of multipath propagation channels is the stochastic tapped-delay-line (STDL) model \cite{hashemi_impulse_1993,cassioli_ultrawide_2002,lee_uwb_2010}, where the path arrival time is quantized into discrete samples whose spacing satisfies the Nyquist criterion. In this model, the delay axis is divided into small time intervals called ``bins". Each bin is assumed to contain either one multipath component or no multipath component. A reasonable bin size is the resolution of the specific measurement system, which is 5 ns in our paper. Using this model, we statistically characterize the received power in each individual time bin, as well as the average power decay pattern. The mathematical expression of the STDL model is as follows
\begin{equation}
	\label{tdl_pt}
	p\left( \tau  \right) = \sum\limits_{k = 0}^{N - 1} {{p_k}\delta \left( {\tau  - {\tau _k}} \right)}, {\tau _k} = k{\tau _s},k \in \mathbb{N},
\end{equation}
where $k$ is the index of the discrete tap, ${p_k}$ is the average power corresponding to each discrete delay ${\tau _k}$, ${\tau _s}$ is the sampling interval, and $N$ is the length of the observation window of the PDP. Considering the spatial scale of the corridor and the laboratory, we choose $N=300$ with a delay span of 1500 ns, corresponding to a physical distance of about 450 m. The PDP characteristics and temporal dispersion parameters within the observation window are investigated.

The PDP may generally decay in accordance with an exponential law \cite{cassioli_ultrawide_2002} or a power law \cite{lee_uwb_2010,ichitsubo_multipath_2000}. The PDP power-law decay satisfies ${p_k} = a/{\tau _k}^n$. Transformed into a logarithmic form, the power-law decay model is expressed as
\begin{equation}
	\label{power-law}
	10{\log _{10}}{p_k} = {\eta _0} - 10{n_{{\rm{PDP}}}}{\log _{10}}{\tau _k} + {X_{{\rm{PDP}}}},
\end{equation}
where $10{\log _{10}}{p_k}$ is linearly related to $10{\log _{10}}{\tau_k}$, $n_{{\rm{PDP}}}$ is the decay factor, ${\eta _0} = 10{\log _{10}}a$ is a constant, and $ {X_{{\rm{PDP}}}} $ is a random variable (in dB) satisfying a normal distribution, i.e. ${X_{{\rm{PDP}}}} \sim \mathcal N \left( {0,{\sigma ^2}} \right)$. Note that the point at $\tau_k = 0$ does not fit with the curve. For a PDP exponential decay, ${p_k} = {e^{ - {\tau _k}/\gamma }}$, its expression in logarithmic form is given by
\begin{equation}
	\label{expo-law}
	10{\log _{10}}{p_k} = \frac{{ - 10{\tau _k}}}{{{\gamma _{{\rm{PDP}}}}\ln 10}} + {X_{{\rm{PDP}}}},
\end{equation}
where the decay factor is $\gamma_{{\rm{PDP}}}$ and the definition of $ {X_{{\rm{PDP}}}} $ remains the same as in the power-law decay model.

In order to explore the attenuation law of PDP, all taps of every PDP are superimposed after power normalization and delay normalization relative to the first tap. Then, the PDP data is fitted according to the PDP model in (\ref{power-law}) and (\ref{expo-law}). The root mean squared error (RMSE) between the actual data $ p_k^\mathrm{meas} $ and the fitted data $ p_k^\mathrm{calc} $ is used to assess the fit of the two models, which is calculated as
\begin{equation}
	\label{rmse}
	{\sigma _{{\rm{RMSE}}}} = \sqrt{\frac{1}{N}\sum\limits_{k = 1}^N {{{\left| {p_k^\mathrm{meas} - p_k^\mathrm{calc}} \right|}^2}} }.
\end{equation}
It is worth noting that $\sigma _{{\rm{RMSE}}}$ and the variance $\sigma$ of $ {X_{{\rm{PDP}}}} $ are numerically equivalent.

\subsection{PDP Analysis}
 
Firstly, based on Figs. \ref{pdp1}-\ref{pdp2}, the PDP characteristics of the three channels at typical measurement points in the corridor and laboratory scenarios are analyzed. Note that the main multipath component is concentrated before 400 ns. As shown in Figs.~\ref{pdp1}(a)-(c), comparing the PDPs of the three measurement points in Row 8 of the corridor scenario, the overall multipath power is lowest in the channel mode without RIS. After the introduction of the RIS, the Tx-RIS-Rx link forms a VLoS path, bringing up the PDP power. In the RIS intelligent reflection mode, the highest tap formed by the VLoS path at each position has a significant power gain, and the power of the PDP is concentrated on the VLoS path. The power level of the RIS specular reflection is between the above two modes.

Figure~\ref{heatmap} depicts the evolution of the PDPs at points in different columns with respect to the RIS-Rx distance, where the color depth represents the power level of the multipath
component. Two noteworthy phenomena can be observed: 1) As the VLoS paths are actively created through RIS phase coding in columns c1 and c3, the channels with RIS intelligent reflection exhibit the best spatial consistency. 2) At the measurement points of column c2 in the corridor (shown as Fig. 6(b) and Fig. 7(b)), the RIS-assisted channels appear to have additional reflected paths with decreasing delay as the RIS-Rx distance increases. This is caused by the multiple reflections between the RIS and the wall and windows at the end of the corridor. However, this phenomenon is not evident in columns c1 and c3.

In the laboratory scenario, as shown in Figs. \ref{pdp2}(a)-(c), the RIS intelligent reflection mode is able to focus the RIS reflected beam onto the Rx and, thus, to maintain the VLoS path power at about -40 dBm, which provides more significant PDP power enhancement compared to the other two modes. It is worth noting that at some points, such as R22, the specularly and intelligently reflected beams of the RIS concentrate the multipath power, but at the same time, lead to the generation of additional multipath components.

\begin{figure}[t]
	\centering
	\includegraphics[width=3in]{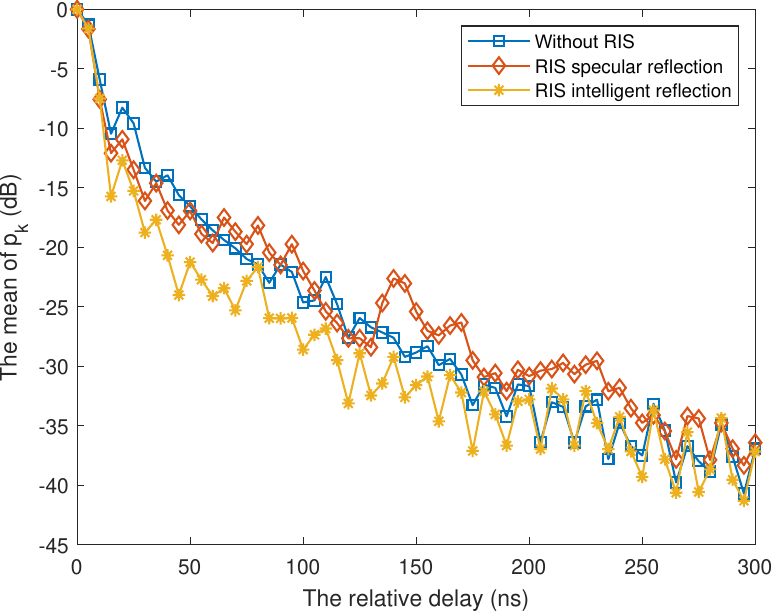}
	\caption{The average PDP decay curves versus the relative delay of the three channel modes in the corridor scenario.}
	\label{pdp_mean}
\end{figure}

In the following, the attenuation pattern of PDPs in the RIS-assisted channel is characterized. After normalizing all the PDPs for the relative power and delay, the power coefficients $p_k$ are averaged over all measurement points. Figure \ref{pdp_mean} plots the average PDP decay curves for the three channels, where the PDP decays significantly before 300 ns and flattens out after 300 ns, implying that the spatial scale of the compacted multipath is about 90 m. In addition, the RIS intelligent reflection mode has the largest attenuation rate of the average power coefficients within the first 100 ns of the PDP in comparison to that of the channel without RIS and with RIS specular reflection.

\begin{figure}[t]
	\centering
	\includegraphics[width=3in]{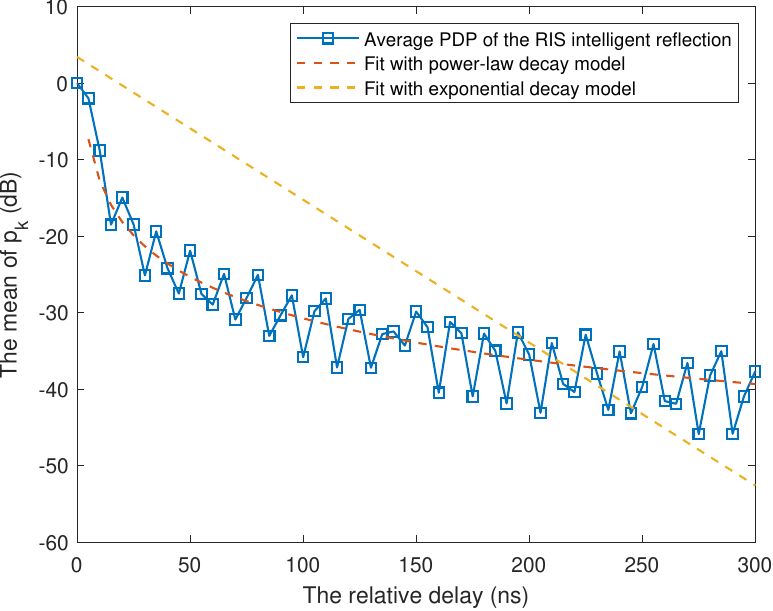}
	\caption{ Comparative curves of the average PDP of the RIS intelligent reflection in the corridor scenario fitting two PDP decay models.}
	\label{pdp_fit}
	\vspace{-0.5cm}
\end{figure}


According to (\ref{power-law}) and (\ref{expo-law}), we use the minimum mean square error (MMSE) method to fit the average PDP curves for the first 300 ns. The fitting parameters are summarized in Table \ref{table2}. It can be seen that the PDPs of the RIS intelligent reflection mode in the corridor and laboratory scenarios fit the power-law decay model better than the exponential decay model and resemble the law of square decay. As shown in Fig. \ref{pdp_fit}, the fitting curve of the exponential decay model cannot keep up with the decay rate of the average PDP. In the corridor scenario, the decay rates of the two models for the RIS specular reflection are slightly smaller than those of the other two channel modes, which may be attributed to the fluctuation of the PDP caused by the secondary reflection paths. In the laboratory scenario, the decay rates of the three channels under the two fading models are characterized by: RIS intelligent reflection $>$ RIS specular reflection $>$ without RIS. Therefore, it can be further deduced that the multipath components are more concentrated around the VLoS path in the RIS intelligent reflection mode, which may alleviate their temporal dispersion.

\begin{table*}[]
	\centering
	\caption{Fitting Parameters of the Two Types of PDP Decay Model\label{table2}}
	\renewcommand{\arraystretch}{1.5}
	\begin{tabular}{|cc|ccc|ccc|}
		\hline
		\multicolumn{2}{|c|}{\multirow{3}{*}{Parameter}} & \multicolumn{3}{c|}{Corridor} & \multicolumn{3}{c|}{Laboratory} \\ \cline{3-8} 
		\multicolumn{2}{|c|}{} & \multicolumn{1}{c|}{Without RIS} & \multicolumn{1}{c|}{\begin{tabular}[c]{@{}c@{}}RIS specular\\ reflection\end{tabular}} & \begin{tabular}[c]{@{}c@{}}RIS intelligent\\ reflection\end{tabular} & \multicolumn{1}{c|}{Without RIS} & \multicolumn{1}{c|}{\begin{tabular}[c]{@{}c@{}}RIS specular\\ reflection\end{tabular}} & \begin{tabular}[c]{@{}c@{}}RIS intelligent\\ reflection\end{tabular} \\ \hline
		\multicolumn{1}{|c|}{\multirow{3}{*}{\begin{tabular}[c]{@{}c@{}}Power-law\\ decay model\end{tabular}}} & ${\eta _0}$ & \multicolumn{1}{c|}{12.78} & \multicolumn{1}{c|}{8.15} & 7.86 & \multicolumn{1}{c|}{13.60} & \multicolumn{1}{c|}{13.07} & 10.44 \\ \cline{2-8} 
		\multicolumn{1}{|c|}{} & $n_{{\rm{PDP}}}$ & \multicolumn{1}{c|}{1.91} & \multicolumn{1}{c|}{1.61} & 1.80 & \multicolumn{1}{c|}{1.86} & \multicolumn{1}{c|}{1.92} & 1.95 \\ \cline{2-8} 
		\multicolumn{1}{|c|}{} & $\sigma _{{\rm{RMSE}}}$ ($\sigma$) & \multicolumn{1}{c|}{6.51} & \multicolumn{1}{c|}{7.00} & 6.02 & \multicolumn{1}{c|}{7.09} & \multicolumn{1}{c|}{6.37} & 6.00 \\ \hline
		\multicolumn{1}{|c|}{\multirow{2}{*}{\begin{tabular}[c]{@{}c@{}}Exponential\\ decay model\end{tabular}}} & $1/{\gamma _{{\rm{PDP}}}}\left( { \times {{10}^{ - 2}}} \right)$ & \multicolumn{1}{c|}{4.07} & \multicolumn{1}{c|}{3.74} & 4.30 & \multicolumn{1}{c|}{3.81} & \multicolumn{1}{c|}{4.03} & 4.39 \\ \cline{2-8} 
		\multicolumn{1}{|c|}{} & $\sigma _{{\rm{RMSE}}}$ ($\sigma$) & \multicolumn{1}{c|}{7.91} & \multicolumn{1}{c|}{7.08} & 9.99 & \multicolumn{1}{c|}{7.31} & \multicolumn{1}{c|}{7.49} & 9.34 \\ \hline
	\end{tabular}
\end{table*}

\subsection{Multipath Parameter Extraction}

In the following, we will identify the multipaths in the PDPs in order to calculate delay dispersion and auxiliary parameters. Considering the multipath features and noise effects, the following multipath identification algorithm is adopted \cite{huang_channel_2021}:

 1) According to (\ref{pt}), find the maximum path power ${P_{\max}}$, take the average of 150 taps in the end part of the PDP as the base noise ${N_{{\rm{noise}}}}$, then, the minimum valid power ${P_{{\rm{min}}}} = \max\left( {{P_{{\rm{max}}}} - 30\:{\rm{dB}}, {N_{{\rm{noise}}}} +   6.6\: {\rm{dB}}} \right)\left( {{\rm{dBm}}} \right)$ is defined as the threshold for screening the non-vanishing multipaths.
 
 2) For index $k=0, 1, ..., N-1$, output the index $k$ and power value ${\rm{PDP}}\left( k \right)$ if ${\rm{PDP}}\left( k \right)$ is greater than the power of both adjacent taps.
 
 3) If ${\rm{PDP}}\left( k \right) < {P_{\min }}$, then discard it. 
 
 4) Record the identified multipath index $l$, multipath delay $\tau_l$, multipath power $P_l$, and calculate the relevant parameters.
 
 \begin{figure}[t]
 	\centering
 	\includegraphics[width=3in]{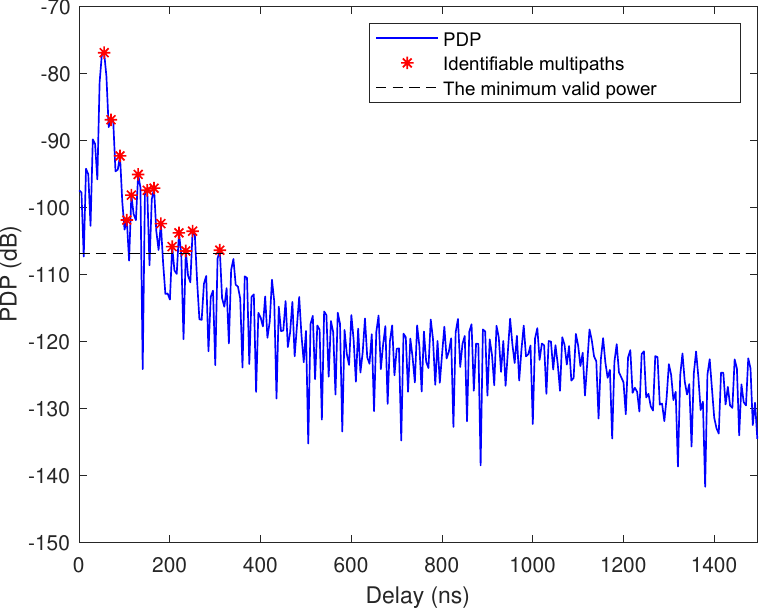}
 	\caption{The example multipath identification result. The minimum valid power ${P_{\min}}$ is calculated according to the maximum path power ${P_{\max}}$ and the base noise ${N_{{\rm{noise}}}}$ of the PDP for screening the non-vanishing multipaths. Red starred dots mark the identified multipath components.}
 	\label{iden-path}
 	\vspace{-0.5cm}
 \end{figure}

The valid multipath threshold $P_{\min}$ is set to be a double-threshold in order to ensure signal integrity by excluding random noise spikes and low signal-to-noise ratio (SNR) artifacts \cite{7996408,huang_channel_2021}, where 30 dB is the minimal dynamic threshold of our system and 6.6 dB is the threshold of the noise floor \cite{Molisch1999}. In addition, for scenarios where LoS paths or VLoS paths are present, the multipath identification starts from the maximum power path, while for NLoS scenarios it starts from an absolute delay of 0 ns. Based on the above steps, the example multipath identification result is shown in Fig. \ref{iden-path}.
We express the identified multipath components as
\begin{equation}
	\label{pl}
	P\left( t \right) = \sum\limits_{l = 1}^L {{P_l}\delta \left( {t - {\tau _l}} \right)},
\end{equation}
where the multipath delay is denoted as $\tau _l$, the multipath power is denoted as $P_l$, and $L$ is the number of multipaths. Notice that unlike the PDP expression, the multipath component appears randomly within a certain delay span.

\begin{table*}[t]
	\centering
	\caption{Multipath Parameters of the Three Channel Modes \label{table3}}
	\renewcommand{\arraystretch}{1.5}
	\begin{tabular}{|c|ccc|ccc|}
		\hline
		\multirow{3}{*}{Parameter} & \multicolumn{3}{c|}{Corridor} & \multicolumn{3}{c|}{Laboratory} \\ \cline{2-7} 
		& \multicolumn{1}{c|}{Without RIS} & \multicolumn{1}{c|}{\begin{tabular}[c]{@{}c@{}}RIS specular\\ reflection\end{tabular}} & \begin{tabular}[c]{@{}c@{}}RIS intelligent\\ reflection\end{tabular} & \multicolumn{1}{c|}{Without RIS} & \multicolumn{1}{c|}{\begin{tabular}[c]{@{}c@{}}RIS specular\\ reflection\end{tabular}} & \begin{tabular}[c]{@{}c@{}}RIS intelligent\\ reflection\end{tabular} \\ \hline
		Multipath power (dB) & \multicolumn{1}{c|}{-51.5} & \multicolumn{1}{c|}{-47.4} & {-41.7} & \multicolumn{1}{c|}{-52.7} & \multicolumn{1}{c|}{-46.9} & {-38.3} \\ \hline
		Maximum path power (dB) & \multicolumn{1}{c|}{-52.9} & \multicolumn{1}{c|}{-48.8} & {-42.4} & \multicolumn{1}{c|}{-54.6} & \multicolumn{1}{c|}{-48.5} & {-39.2} \\ \hline
		Maximum path delay (ns) & \multicolumn{1}{c|}{51.5} & \multicolumn{1}{c|}{47.5} & {46.0} & \multicolumn{1}{c|}{48.8} & \multicolumn{1}{c|}{46.5} & {43.4} \\ \hline
		Mean delay (ns) & \multicolumn{1}{c|} {58.6} & \multicolumn{1}{c|}{59.3} & {47.8} & \multicolumn{1}{c|}{61.4} & \multicolumn{1}{c|}{54.3} & {45.0} \\ \hline
		RMS delay spread (ns) & \multicolumn{1}{c|}{25.9} & \multicolumn{1}{c|}{34.4} & {21.7} & \multicolumn{1}{c|}{31.3} & \multicolumn{1}{c|}{25.7} & {18.5} \\ \hline
		K-factor (dB) & \multicolumn{1}{c|} {4.5} & \multicolumn{1}{c|}{6.4} & {12.5} & \multicolumn{1}{c|}{7.3} & \multicolumn{1}{c|}{6.3} & {12.3}\\ \hline
	\end{tabular}
\end{table*}

Based on the multipath delay and power obtained above, the extraction methods for the parameters, such as received power, mean delay, and root mean squared  (RMS) delay spread will be presented below, respectively. Firstly, the received power, which is the total power of the identified multipaths, is calculated as
\begin{equation}
	\label{pr}
	{P_r} = \sum\limits_{l = 1}^L {P_l}.
\end{equation}
The mean delay and the RMS delay spread are the two main temporal dispersion parameters. The mean delay represents the dispersion characteristics of different paths in the delay domain and is numerically expressed as a power-weighted average delay of multipaths as follows:
\begin{equation}
	\label{tmean}
	\bar \tau  = \frac{{\sum\limits_{l = 1}^L {{\tau _l}{P_l}} }}{{\sum\limits_{l = 1}^L {P_l} }}.
\end{equation}
The RMS delay spread is inversely related to the coherence bandwidth and is one of the key parameters in the design of the physical layer of wireless communications. Its mathematical expression corresponds to the power-weighted variance of the delay of the multipath component, which is
\begin{equation}
	\label{trms}
	{\tau _{{\rm{RMS}}}} = \sqrt {\frac{{\sum\limits_{l = 1}^L {{{\left( {{\tau _l} - \bar \tau } \right)}^2}} {P_l}}}{{\sum\limits_{l = 1}^L {P_l}}}}.
\end{equation}

In addition, the Ricean K factor represents the ratio between the deterministic component power and the stochastic component power in a Ricean channel. To calculate a wideband K-factor, a sub-band technique as described in Eqs. (8)-(10) of \cite{tang_estimation_2019} is utilized.

\subsection{Multipath Parameter Analysis}
The multipath components in the PDPs of the three channel modes are extracted and the relevant parameters are calculated. Their mean values are summarized in Table \ref{table3} and analyzed as follows. In the corridor scenario, the RIS intelligent reflection improves the received power by an average of about 6 dB over the RIS specular reflection and about 10 dB over without RIS. In the laboratory scenario, the RIS intelligent reflection improves the received power by an average of about 8.5 dB over the RIS specular reflection and about 14 dB over without RIS. The power improvement is mainly due to the increase in the maximum path power.

The K-factor of the RIS intelligent reflection in the corridor scenario is 8 dB more than that in the channel without RIS. In the laboratory scenario, the K-factor of the RIS intelligent reflection increases 5 dB compared to the channel without RIS. The above results indicate that the RIS intelligent reflection improves significantly the K-factor and concentrates the deterministic component of the channel.

In addition to the power variations, the RIS with intelligent reflection changes the multipath delay. In the corridor scenario, the RIS with intelligent reflection reduces the maximum path delay by about 5.5 ns and the mean delay by about 10.8 ns on average compared to without RIS. In the laboratory scenario, the mean delay of the RIS intelligent reflection decreases about 16 ns compared to without RIS.

\begin{figure*}[t]
	\centering
	\subfloat[]{\includegraphics[width=3in]{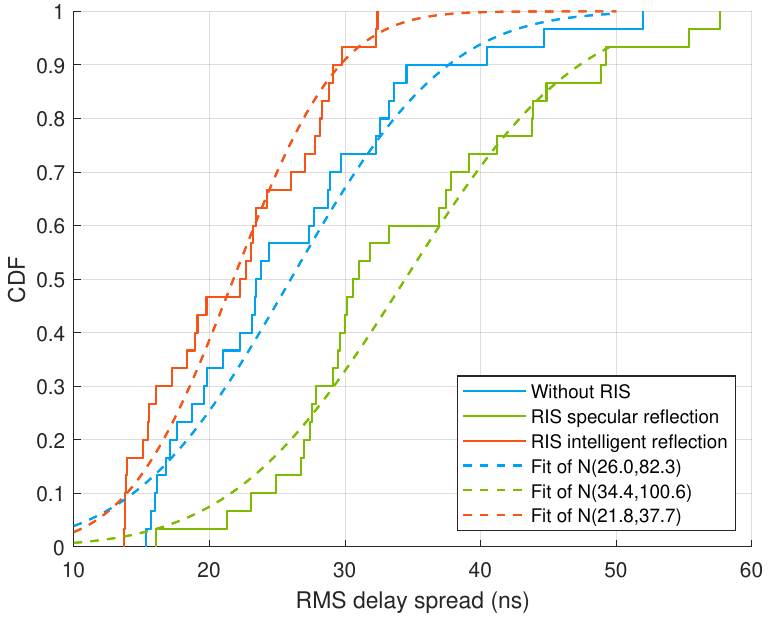}}
	\hfil
	\subfloat[]{\includegraphics[width=3in]{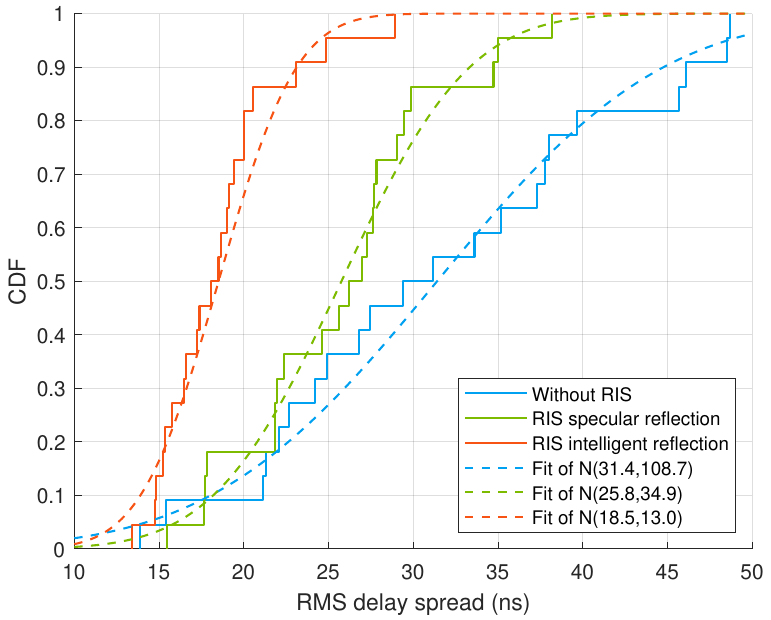}}
	\caption{CDFs of the RMS delay spread for three channel modes in different scenarios. The measured CDFs and their Gaussian distribution fitted CDFs noted as $N(\mu, \sigma^2)$ are depicted as solid and dashed lines, respectively. (a) CDFs of the RMS delay spread in the corridor scenario. (b) CDFs of the RMS delay spread in the laboratory scenario.}
	\label{cdf}
	\vspace{-0.3cm}
\end{figure*}

Furthermore, the cumulative distribution function (CDF) of the RIS delay spread in the three channel modes and its Gaussian distribution fitted CDF are shown in Fig. \ref{cdf}. Compared with the conventional wireless channel without RIS, the RMS delay spread of the corridor scenario with RIS intelligent reflection is reduced by about 4.2 ns on average. This is because the RIS intelligent reflection reduces the maximum path delay and concentrates the maximum path power. However, the delay spread of the RIS specular reflection is increased by about 8.5 ns, which is because reflected clusters generated between the RIS and the wall cause multipath extension in the delay domain. In the laboratory scenario, the RMS delay spread of the RIS specular reflection is reduced by about 5.6 ns on average, and the reduction of the RIS intelligent reflection is about 13 ns on average. In conclusion, through intelligent reflection based on the configuration method in Eq. (\ref{phi}), the RIS is able to eliminate the phase differences of the multipaths reflected by the RIS and concentrate the multipath energy. Hence, the time-domain multipath dispersion effect is remarkably alleviated.

It can be noted that the power enhancement and multipath fading alleviation introduced by the RIS-assisted channel in the laboratory environment is more substantial in comparison to that in the corridor scenario, suggesting that the wireless channel environment is an important factor in the process of active manipulation of the wireless channel propagation through RIS. In narrow spaces, such as corridors, a RIS contributes to the creation of additional reflection paths, while in more open laboratories, a RIS reduces the delay spread and mitigates multipath fading more effectively.

\section{Characteristics of Multipath Clustering in RIS-assisted Channels}

\subsection{Cluster Model}
In the propagation process of wireless signals, radio waves are reflected, scattered, and diffracted by objects in the environment, producing many multipath components propagating along different paths with different amplitudes, phases, and angles of arrival. Some multipath components experience a similar reflection process in the wireless channel, and the variance in the delay and angle of arrival between them is small. Thus, a group of multipath components with similar parameters is classified as a cluster. The S-V model \cite{saleh_statistical_1987} is a commonly used clustering model that can be widely used in indoor and outdoor environments. In this paper, the multipath clustering characteristics of RIS-assisted channels are modeled using the S-V model. The S-V model expresses the CIR as
\begin{equation}
	\label{sv}
	h(\tau ) = \sum\limits_{l = 0}^{L_c} {\sum\limits_{k = 0}^{{K_l}} {{\beta _{kl}}{e^{j{\theta _{kl}}}}} } \delta \left( {\tau  - {T_l} - {\tau _{kl}}}\right),
\end{equation}
where $L_c$ is the number of clusters and $K_l$ is the number of rays in the $l$-th cluster. The parameter $ T_l $ denotes the arrival delay of the $l$-th cluster, ${\tau _{kl}}$ is the relative delay of the $k$-th ray in the $l$-th cluster, while the phase of ray $\theta _{kl}$ obeys the uniform distribution. The amplitude of the $k$-th ray in the $l$-th cluster is denoted as $\beta _{kl}$, while its average power satisfies the following decay pattern
\begin{equation}
	\label{betae}
	\overline {\beta _{kl}^2}  = \overline {\beta _{00}^2} {e^{ - {T_l}/\Gamma }}{e^{ - {\tau _{kl}}/\gamma }},
\end{equation}
where the cluster decay factor $1/\Gamma$ and the ray decay factor $1/\gamma$ characterize the average attenuation rate of the overall envelope of the cluster and the ray envelope within the cluster, respectively.

Typically, the power level (in dB) of each successive multipath component is linearly correlated with its delay, which leads to the exponential decay rule of multipath power shown in (\ref{betae}) \cite{saleh_statistical_1987}. However, in LoS scenarios where strong reflections exist, a power-law decay model was found to be more suitable for measurement data \cite{karedal_measurementbased_2007,ai_power_2015}, where the multipath log-value power decays logarithmically with the delay, i.e.
\begin{equation}
	10{\log _{10}}\overline {\beta _{kl}^2} = {\eta _0} - 10{n_{\rm ray}}{\log _{10}}{\tau _{kl}},\label{betan}
\end{equation}
where ${\eta _0}$ is a constant and $n_{\rm ray}$ denotes the average power decay factor for rays. The same equation in the above model can be applied to the cluster power $\overline {\beta _{0l}^2}$, cluster delay $T _{l}$, and the decay factor $n_{\rm cluster}$. In RIS-assisted channels, the RIS cascaded channel forms an extremely strong VLoS path and strong secondary reflected paths, making the power-law decay model more applicable. 

In addition, the main parameters describing the characteristics of multipath clustering are the number of clusters, inter-cluster interval, and intra-cluster RMS delay spread. In the S-V model, the number of clusters is modeled using a Poisson distribution \cite{meijerink_physical_2014}:
\begin{equation}
	\label{N}
	\begin{aligned}
		N &\sim {\rm{P}} \left( {\lambda {T_{\rm dur}}} \right)\\
		P(n = N) &= \frac{{{{\left( {\lambda {T_{\rm dur}}} \right)}^n}{e^{ - \lambda {T_{\rm dur}}}}}}{{n!}},
	\end{aligned}
\end{equation}
where $T_{\rm dur}$ is the total delay span of the observation window, $\lambda$ is the average arrival rate of the clusters, which is equal to the quotient of the average number of clusters and the total delay $T_{\rm dur}$, while the number of clusters is obtained by the clustering algorithm. The inter-cluster interval $T_l$ fulfills the exponential distribution
\begin{equation}
	\label{pT}
	P(T_l) = \lambda {e^{ - \lambda \left( {T_l - {\tau _0}} \right)}},
\end{equation}
where ${\tau _0}$ is an offset which is equal to the minimum value of the inter-cluster interval \cite{ai_power_2015}. The intra-cluster RMS delay spread is an important parameter in the design of protection intervals for communication systems, which can be modeled as a log-normal distributed variable $10\rm log_{10}$$ \tau _l^{\rm {RMS}} \sim \mathcal N (\tau_l,\sigma_l^2)$, with the expression
\begin{equation}
	\label{tl}
	{\tau _l^{\rm RMS}} = \sqrt {\frac{{\sum\limits_{k = 0}^{{K_l}} {{{\left( {{\tau _{kl}} - {\tau _{{\rm{mean}}}}} \right)}^2}} \overline {\beta _{kl}^2} }}{{\sum\limits_{k = 0}^{{K_l}} {\overline {\beta _{kl}^2} } }}},
\end{equation}
while the mean intra-cluster delay is
\begin{equation}
	\label{tm}
	{\tau _{{\rm{mean}}}} = \frac{{\sum\limits_{k = 0}^{{K_l}} {{\tau _{kl}}} \overline {\beta _{kl}^2} }}{{\sum\limits_{k = 0}^{{K_l}} {\overline {\beta _{kl}^2} } }}.
\end{equation}

\subsection{Clustering Algorithm}

In order to extract the above multipath cluster parameters, the multipath components identified in Section \Rmnum{3}-C need to be clustered at first. In the measurements of the RIS-assisted channel in this paper, the multipath components have two characteristic parameters, namely delay and power. Due to the small number of multipaths, the clustering results based on the K-PowerMeans algorithm \cite{czink_framework_2006} are not satisfactory. Furthermore, during the clustering process we face the problems of determining the number of clusters and selecting the proper initial cluster center. 

\begin{algorithm}[t]
	\renewcommand{\algorithmicrequire}{\textbf{Input:}}
	\renewcommand{\algorithmicensure}{\textbf{Output:}}
	\caption{The improved S-V model bubbling clustering algorithm}
	\begin{algorithmic}[1]
		\REQUIRE $\{PDP(\tau)\}$, $T_0$, $T_{\rm{dur}}$, $T_{\rm{dis}}$, $P_{\rm{dis}}$, $\beta _{\rm{offset}}^2$
		\ENSURE $\{PDP_{\rm{en}}(\tau_{\rm{en}})\}$, $\{PDP_{\rm{lo}}(\tau_{\rm{lo}})\}$, $\{\tau_{\rm{out}}\}$
		\STATE Search for local maximums of $\{PDP(\tau)\}$ within the delay $[T_0, T_0+T_{\rm{dur}}]$ and record results as the multipath collection $\{PDP_{\rm{en}}(\tau_{\rm{en}})\}$
		\STATE Search for local maxima in $\{PDP_{\rm{en}}(\tau_{\rm{en}})\}$ with a search step of $T_{\rm{dis}}$ and the minimal peak prominence of $P_{\rm{dis}}$, record results as the local maximum collection $\{PDP_{\rm{lo}}(\tau_{\rm{lo}})\}$ 
		\IF{The maximal entry $PDP_{\rm{max}}(\tau_{\rm{max}})$ of $\{PDP(\tau)\}$ is the first element in $\{PDP_{\rm{en}}(\tau_{\rm{en}})\}$} 
		\STATE $\{PDP_{\rm{lo}}(\tau_{\rm{lo}})\} \gets \{ PDP_{\rm{max}}(\tau_{\rm{max}}), PDP_{\rm{lo}}(\tau_{\rm{lo}})\}$ 
		\ENDIF 
		\STATE Define $i \gets size(\{PDP_{\rm{lo}}\})$ 
		\WHILE{$i \neq 0$} 
		\STATE Search for the maximum of $\{PDP(\tau)\}$ within the delay $[\tau_{\rm{lo}}(i), T_{\rm{dur}}]$ and record the power as $\beta _{\rm{max}}^2$
		\IF{$PDP_{\rm{lo}}(i) \geq \beta _{\rm{max}}^2 - \beta _{\rm{offset}}^2$} 
		\STATE $\{\tau_{\rm{out}}\}$ $\gets$ $\{ \tau_{\rm{out}}, \tau_{\rm{lo}}(i)\}$
		\ENDIF 
		\STATE $i \gets i - 1$ 
		\ENDWHILE 
	\end{algorithmic}
	\label{alg1}
\end{algorithm}

\begin{table}[t]
	\centering
	\caption{Parameter Definitions and Values of Algorithm 1\label{table5}}
	\renewcommand{\arraystretch}{1.5}
	\begin{tabular}{|c|c|c|}
		\hline
		\textbf{Parameter} & \textbf{Symbol} & \textbf{Value} \\ \hline
		PDP snapshot & $\{PDP(\tau)\}$ & - \\ \hline
		Initial delay & $T_0$ & 0 ns \\ \hline
		Delay span & $T_{\rm{dur}}$ & 300 ns \\ \hline
		Delay resolution & $\Delta \tau$ & 5 ns \\ \hline
		Search step & $T_{\rm{dis}}$ & 30 ns \\ \hline
		Minimal peak prominence & $P_{\rm{dis}}$ & 5 ns \\ \hline
		Power offset threshold & $\beta _{\rm{offset}}^2$ & 1 dB \\ \hline
		\begin{tabular}[c]{@{}c@{}}Set of multipath envelope\\ powers and delay pairs\end{tabular} & $\{PDP_{\rm{en}}(\tau_{\rm{en}})\}$ & - \\ \hline
		\begin{tabular}[c]{@{}c@{}}Set of local maximal multipath\\ powers and delay pairs\end{tabular} & $\{PDP_{\rm{lo}}(\tau_{\rm{lo}})\}$ & - \\ \hline
		\begin{tabular}[c]{@{}c@{}}Set of the first arrival\\ ray delays in clusters\end{tabular} & $\{\tau_{\rm{out}}\}$ & - \\ \hline
	\end{tabular}
\end{table}

\begin{figure}[h]
	\centering
	\includegraphics[width=3in]{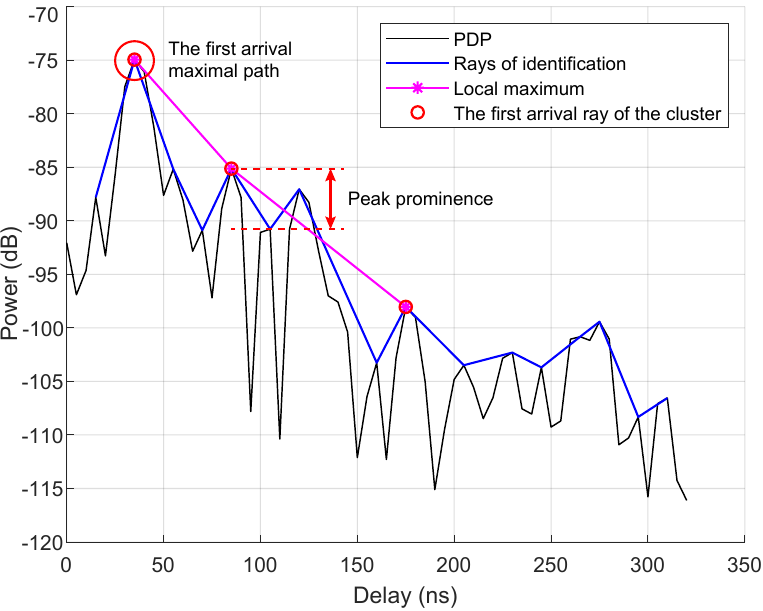}
	\caption{The illustration of the clustering process using the improved algorithm for an example PDP. First, identify the rays of the PDP, then search for the local maxima of rays with a certain searching step, and finally find the local maxima that satisfies the monotonous decay and the peak prominence threshold as the first arriving path of the cluster. Add and label the first arrival maximal path as the first ray in the initial cluster. }
	\label{S-V-clstering}
\end{figure}

\begin{table*}[t]
	\centering
	\caption{Cluster Parameters of The Three Channel Modes\label{table4}}
	\renewcommand{\arraystretch}{1.5}
	\begin{tabular}{|c|c|ccc|ccc|}
		\hline
		\multirow{3}{*}{Parameter} & Scenario & \multicolumn{3}{c|}{Corridor} & \multicolumn{3}{c|}{Laboratory} \\ \cline{2-8} 
		& Channel mode & \multicolumn{1}{c|}{Without RIS} & \multicolumn{1}{c|}{\begin{tabular}[c]{@{}c@{}}RIS specular\\ reflection\end{tabular}} & \begin{tabular}[c]{@{}c@{}}RIS intelligent\\ reflection\end{tabular} & \multicolumn{1}{c|}{Without RIS} & \multicolumn{1}{c|}{\begin{tabular}[c]{@{}c@{}}RIS specular\\ reflection\end{tabular}} & \begin{tabular}[c]{@{}c@{}}RIS intelligent\\ reflection\end{tabular} \\ \hline
		\begin{tabular}[c]{@{}c@{}} Average number \\ of clusters \end{tabular}& $L_c$ & \multicolumn{1}{c|}{1.43} & \multicolumn{1}{c|}{2.56} & 2.13 & \multicolumn{1}{c|}{1.54} & \multicolumn{1}{c|}{1.54} & 1.45 \\ \hline
		\begin{tabular}[c]{@{}c@{}}Inter-cluster\\ interval\end{tabular} & \begin{tabular}[c]{@{}c@{}}${T_l}$ $(\rm {ns})$\end{tabular} & \multicolumn{1}{c|}{235.25} & \multicolumn{1}{c|}{130.12} & 163.70 & \multicolumn{1}{c|}{207.04} & \multicolumn{1}{c|}{220.79} & 222.38 \\ \hline
		\multirow{2}{*}{\begin{tabular}[c]{@{}c@{}}Intra-cluster\\ RMS delay spread\end{tabular}} & \begin{tabular}[c]{@{}c@{}}${\tau _l}$ $ (\rm log_{10} (\rm ns))$ \end{tabular} & \multicolumn{1}{c|}{3.18} & \multicolumn{1}{c|}{3.02} & 2.94 & \multicolumn{1}{c|}{3.26} & \multicolumn{1}{c|}{3.16} & 3.09 \\ \cline{2-8} 
		& ${\sigma _l}$ & \multicolumn{1}{c|}{0.35} & \multicolumn{1}{c|}{0.62} & 0.54 & \multicolumn{1}{c|}{0.38} & \multicolumn{1}{c|}{0.37} & 0.35 \\ \hline
		\begin{tabular}[c]{@{}c@{}}Arrival rate\\ of clusters\end{tabular} & \begin{tabular}[c]{@{}c@{}}$\lambda $ $(×10^{-3})$\end{tabular} & \multicolumn{1}{c|}{4.8} & \multicolumn{1}{c|}{8.6} & 7.1 & \multicolumn{1}{c|}{5.2} & \multicolumn{1}{c|}{5.2} & 4.8 \\ \hline
		\begin{tabular}[c]{@{}c@{}}Decay factor\\ of clusters\end{tabular} & ${n_{{\rm{cluster}}}}$ & \multicolumn{1}{c|}{1.05} & \multicolumn{1}{c|}{0.77} & 1.00 & \multicolumn{1}{c|}{0.80} & \multicolumn{1}{c|}{0.95} & 1.03 \\ \hline
		\begin{tabular}[c]{@{}c@{}}Decay factor\\ of rays\end{tabular} & ${n_{{\rm{ray}}}}$ & \multicolumn{1}{c|}{1.06} & \multicolumn{1}{c|}{0.75} & 0.88 & \multicolumn{1}{c|}{0.96} & \multicolumn{1}{c|}{1.02} & 1.14 \\ \hline
		\begin{tabular}[c]{@{}c@{}}Decay factor of rays\\ in the first cluster\end{tabular} & ${n_{{\rm{first-cluster}}}}$ & \multicolumn{1}{c|}{1.29} & \multicolumn{1}{c|}{0.88} & 1.23 & \multicolumn{1}{c|}{1.11} & \multicolumn{1}{c|}{1.25} & 1.38 \\ \hline
	\end{tabular}
\end{table*}

In addition to traditional clustering algorithms, some heuristic algorithms \cite{lee_uwb_2010,woon_identification_2006,jiang_comparative_2020} have been proposed for cluster identification scenarios with only multipath power and delay as feature vectors. However, it is very difficult to design a robust estimation algorithm for different scenarios. In this paper, by leveraging the actual characteristics of RIS-assisted channels, the ``bubbling" clustering algorithm, based on the S-V model proposed in \cite{jiang_comparative_2020}, is selected and improved to meet our practical applications. The ``bubbling" algorithm is based on the S-V model and follows two criteria \cite{jiang_comparative_2020}: 1) The power of the first arrival ray in a cluster is a local maximum in the PDP. 2) The power of the first arrival ray in each cluster is monotonically decreasing at any time delay. Its main idea is to search for the peak of the PDP (equivalent to identifying multipaths), then search for the local maxima across multipaths with a certain searching step, and finally find the local peak that satisfies the monotonous decay as the first arriving path of the cluster from back to front. The advantage of this method is that it contains the multipath identification process, which ensures the integrity of the multipath components. Moreover, it shows the multipath evolution in the process of clustering, which is beneficial to the analysis of cluster parameters and avoids the selection of the number of clusters and the initial cluster center. 

Nevertheless, clustering using the above S-V model bubbling algorithm may suffer from the following problems:

1) The secondary search for local peaks may miss identifying the first ray when it is the maximum path;

2) When the peak of a chosen first arriving ray is not dominant, it is more proper to view that ray as a random intra-cluster fluctuation or trailing of the previous cluster rather than as the first arrival ray of a new cluster.

To address these issues, we improve the S-V model bubbling clustering algorithm by adding the first reaching maximum path and introducing the minimum peak prominence parameter, so that the clustering results can be adjusted according to the actual measurement data. The improved S-V model bubbling clustering algorithm is summarized in Algorithm \ref{alg1} and its main parameter definitions and values are presented in Table \ref{table5}. Using the improved algorithm, Fig. \ref{S-V-clstering} shows a schematic diagram of the key steps of clustering a PDP at a certain measurement location, where two improvement points are marked.

\begin{figure*}[t]
	\centering
	\subfloat[]{\includegraphics[width=2.4in]{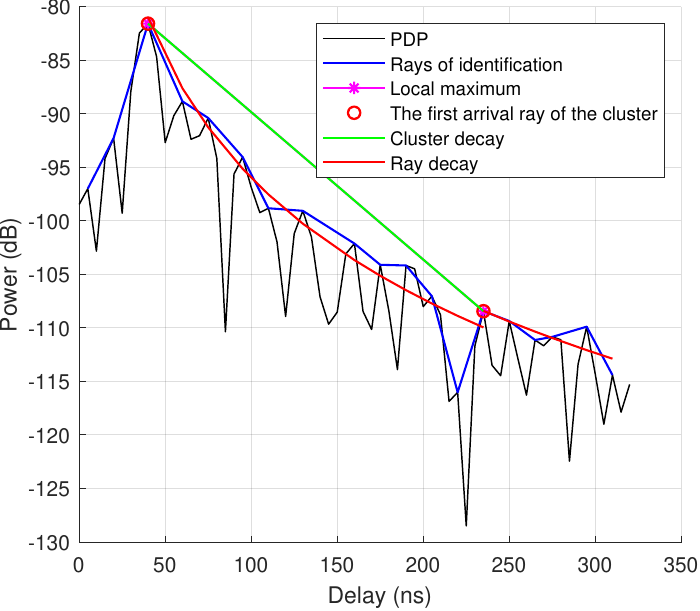}}
	\subfloat[]{\includegraphics[width=2.4in]{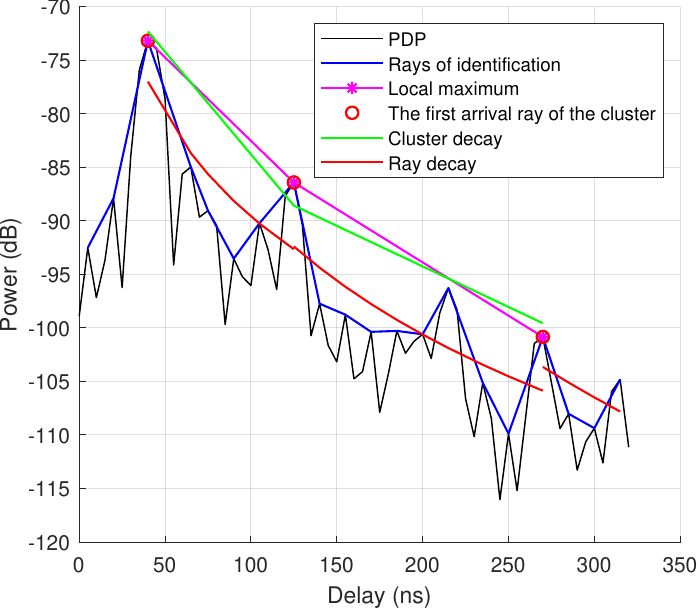}}
	\subfloat[]{\includegraphics[width=2.4in]{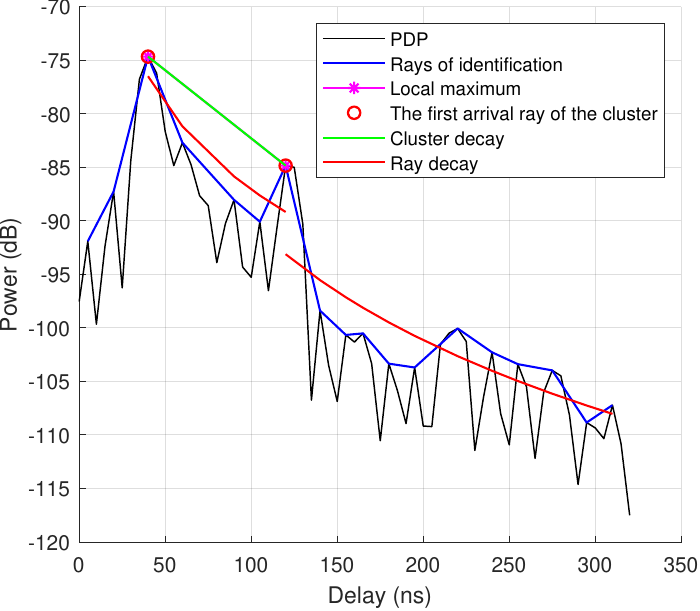}}
	\\
	\subfloat[]{\includegraphics[width=2.4in]{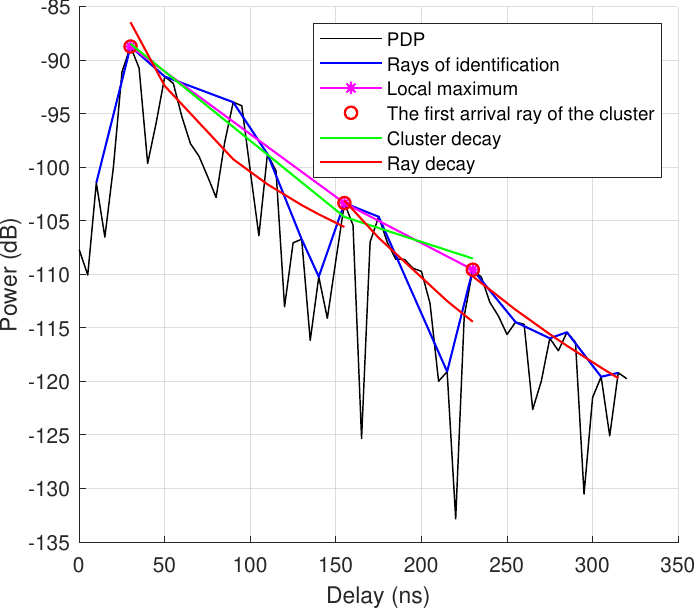}}
	\subfloat[]{\includegraphics[width=2.4in]{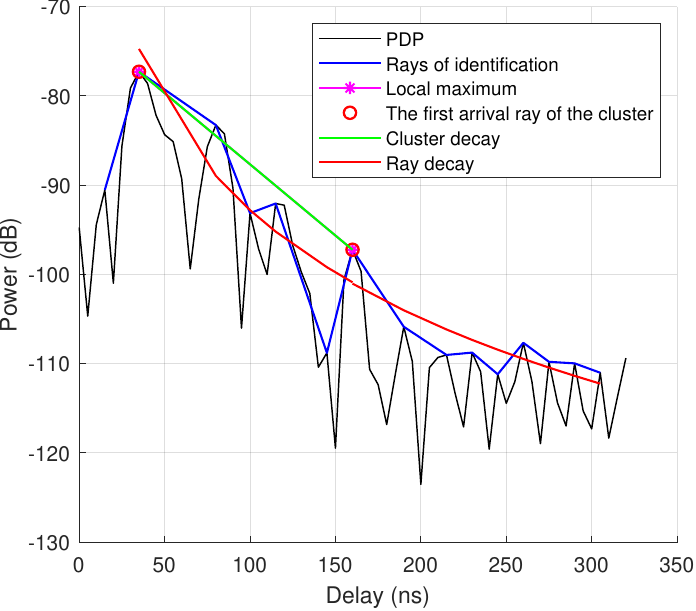}}
	\subfloat[]{\includegraphics[width=2.4in]{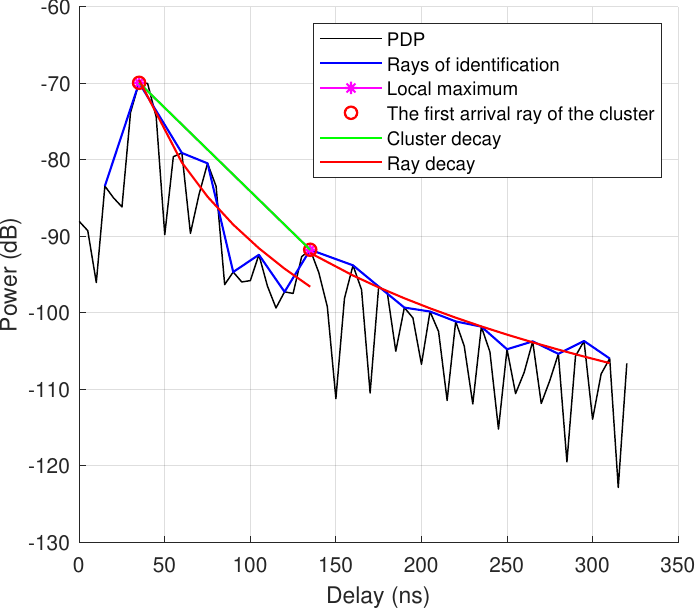}}
	\caption{Multipath clustering results for three channel modes at typical measurement points. The fits of cluster decay and ray decay are represented by the green curve and the red curve using the power-law decay model. (a) Clustering result at Row 5, Column 2 of the channel without RIS in the corridor scenario. (b) Clustering result at Row 5, Column 2 of the channel with the RIS specular reflection in the corridor scenario. (c) Clustering result at Row 5, Column 2 of the channel with the RIS intelligent reflection in the corridor scenario. (d) Clustering result at R5 of the channel without RIS in the laboratory scenario. (e) Clustering result at R5 of the channel with the RIS specular reflection in the laboratory scenario. (f) Clustering result at R5 of the channel with the RIS intelligent reflection in the laboratory scenario.}
	\label{cluster1}
\end{figure*}

\subsection{Cluster Characteristics}
Based on the improved S-V model bubbling clustering algorithm, the multipath components of PDPs within the delay range from 0 ns to 300 ns for the three channel modes are clustered. The cluster parameters are analyzed in the following. We extract the average decay factors based on the power-law decay model in (\ref{betan}). According to Section \Rmnum{4}-A, the average number of clusters, intra-cluster parameters, and inter-cluster parameters in the corridor and laboratory scenarios are summarized in Table \ref{table4}. In order to explain the above parameters more explicitly, Fig. \ref{cluster1} illustrates the multipath clustering results and plots the fitting curves of the cluster decay as well as ray decay for three channel modes at typical measurement points.

In the corridor scenario, the average number of clusters reflects the fact that there are more clusters in RIS-assisted channels. As shown in Fig. \ref{cluster1}(b) and \ref{cluster1}(c), multiple reflected paths promote the formation of more clusters due to the waveguide-like spatial structure of the corridor scenario. Specifically, after placing the RIS at the corridor turn, the transmitting signal is reflected by the RIS to form a strong beam, and then the reflected beam is reflected back and forth between the walls around the corridor or reflected by the metal window frames at the end of the corridor, generating clusters at different delays. Conversely, there does not exist a VLoS path in the channel without RIS to produce multiple strong reflections. As shown in Fig. \ref{cluster1}(b), the RIS specular reflection produces a wide reflected beam and, thus, the ``multiple reflected clusters" phenomenon is more obvious. The RIS intelligent reflection mitigates this phenomenon by beam focusing, but it introduces additional clusters compared to the case without RIS. Consequently, the average number of clusters is the largest for the RIS specular reflection and smallest for the mode without RIS in the corridor scenario.

An increase in the number of clusters leads to a decrease in the inter-cluster interval, whose mean values satisfy: without RIS $>$ RIS intelligent reflection $>$ RIS specular reflection. Recall that the cluster arrival rate is equal to the ratio of the number of clusters to the length of the delay span. Therefore, the cluster arrival rate exhibits a magnitude relationship of: RIS specular reflection $>$ RIS intelligent reflection $>$ without RIS. In order to provide a reference for analyzing the decay factors, we list the ray decay factor for the first cluster in Table \ref{table4} along with the average cluster decay factor and the average ray decay factor for all clusters. In the case of the RIS intelligent reflection, it is worth noting that the average decay factor of rays is 0.88, while the decay factor of rays in the first cluster reaches 1.23, which reveals that the ray attenuation occurs mainly within the first cluster and slows down substantially in the later clusters. Furthermore, the additional reflected clusters cause the reduction of the cluster decay factors in RIS-assisted channels.

In the laboratory scenario, the average number of clusters of the RIS intelligent reflection decreases relative to the channel without RIS, with a consequent decrease in the cluster arrival rate. At the same time, the cluster decay factor of the RIS intelligent reflection turns to be the largest one, which means that the RIS intelligent reflection concentrates the multipath energy on the VLoS path. Figures \ref{cluster1}(d)-(f) show the multipath clustering results for three channel modes at measurement point R5 in the laboratory. Affected by the scatterers in the laboratory, the received signal without the RIS's assistance has strong multipath fading characteristics. By contrast, the channels with RIS intelligent reflection always present a first strong cluster dominated by VLoS path, whereas the power of other clusters is suppressed and the multipath envelope is smoothly attenuated.

\section{Conclusion}
In RIS-assisted wireless communication systems, channel measurements and modeling are essential for the system design, network optimization and performance evaluation. In order to explore the small-scale fading characteristics of RIS-assisted channels, this paper built a RIS time-domain channel measurement system based on USRP. We compared the PDP, multipath parameters, and clustering parameters of the channel without RIS, with RIS specular reflection, and with RIS intelligent reflection in corridor and laboratory scenarios, respectively. The PDPs of the RIS intelligent reflection mode in the corridor and laboratory scenarios tend to better fit the power-law decay model rather than the exponential decay model and resemble the law of square decay. Furthermore, the multipath and cluster characteristics of the RIS-assisted channel were analyzed. Measurement results demonstrated that a RIS with intelligent reflection can enhance and concentrate the energy of the VLoS path reflected by the RIS and, thus, mitigate multipath fading and reduce delay spread. Regarding the cluster characteristics, a single cluster dominated by a VLoS path with smooth envelope was observed in the laboratory scenario. In the corridor scenario, the additional reflected clusters of the RIS specular reflection channel and the RIS intelligent reflection channel were presented, where the RIS undertook the role of  generating multiple reflected paths.

\bibliographystyle{IEEEtran}
\bibliography{RIS_channel.bib}

\begin{thebibliography}{10}
\providecommand{\url}[1]{#1}
\csname url@samestyle\endcsname
\providecommand{\newblock}{\relax}
\providecommand{\bibinfo}[2]{#2}
\providecommand{\BIBentrySTDinterwordspacing}{\spaceskip=0pt\relax}
\providecommand{\BIBentryALTinterwordstretchfactor}{4}
\providecommand{\BIBentryALTinterwordspacing}{\spaceskip=\fontdimen2\font plus
\BIBentryALTinterwordstretchfactor\fontdimen3\font minus
  \fontdimen4\font\relax}
\providecommand{\BIBforeignlanguage}[2]{{%
\expandafter\ifx\csname l@#1\endcsname\relax
\typeout{** WARNING: IEEEtran.bst: No hyphenation pattern has been}%
\typeout{** loaded for the language `#1'. Using the pattern for}%
\typeout{** the default language instead.}%
\else
\language=\csname l@#1\endcsname
\fi
#2}}
\providecommand{\BIBdecl}{\relax}
\BIBdecl

\bibitem{shafi_5g_2017}
M.~Shafi \emph{et~al.}, ``{{5G}}: {{A}} tutorial overview of standards, trials,
  challenges, deployment, and practice,'' \emph{IEEE J. Sel. Areas Commun.},
  vol.~35, no.~6, pp. 1201--1221, Jun. 2017.

\bibitem{Matthaiou:COMMag:2021}
M.~Matthaiou, O.~Yurduseven, H.~Q. Ngo, D.~Morales-Jimenez, S.~L. Cotton, and
  V.~F. Fusco, ``The road to {6G: Ten} physical layer challenges for
  communications engineers,'' \emph{IEEE Commun. Mag.}, vol.~59, no.~1, pp.
  64--69, Jan. 2021.

\bibitem{zhang2020prospective}
J.~Zhang, E.~Bj{\"o}rnson, M.~Matthaiou, D.~W.~K. Ng, H.~Yang, and D.~J. Love,
  ``Prospective multiple antenna technologies for beyond {5G},'' \emph{IEEE J.
  Sel. Areas Commun.}, vol.~38, no.~8, pp. 1637--1660, Aug. 2020.

\bibitem{chen_survey_2019}
Z.~Chen \emph{et~al.}, ``A survey on terahertz communications,'' \emph{China
  Commun.}, vol.~16, no.~2, pp. 1--35, Feb. 2019.

\bibitem{han_large_2019}
Y.~Han, W.~Tang, S.~Jin, C.-K. Wen, and X.~Ma, ``Large intelligent
  surface-assisted wireless communication exploiting statistical {{CSI}},''
  \emph{IEEE Trans. Veh. Technol.}, vol.~68, no.~8, pp. 8238--8242, Aug. 2019.

\bibitem{9340586}
Y.~Lin, S.~Jin, M.~Matthaiou, and X.~You, ``Tensor-based algebraic channel
  estimation for hybrid {IRS}-assisted {MIMO-OFDM},'' \emph{IEEE Trans.
  Wireless Commun.}, vol.~20, no.~6, pp. 3770--3784, Jun. 2021.

\bibitem{direnzo_smart_2020}
M.~Di~Renzo \emph{et~al.}, ``Smart radio environments empowered by
  reconfigurable intelligent surfaces: {{How}} it works, state of research, and
  the road ahead,'' \emph{IEEE J. Sel. Areas Commun.}, vol.~38, no.~11, pp.
  2450--2525, Nov. 2020.

\bibitem{wu_intelligent_2021}
Q.~Wu, S.~Zhang, B.~Zheng, C.~You, and R.~Zhang, ``Intelligent reflecting
  surface-aided wireless communications: {{A}} tutorial,'' \emph{IEEE Trans.
  Commun.}, vol.~69, no.~5, pp. 3313--3351, May 2021.

\bibitem{sang_coverage_2022}
J.~\vspace{0mm}Sang \emph{et~al.}, ``Coverage enhancement by deploying {{RIS}}
  in {{5G}} commercial mobile networks: {{Field}} trials,'' \emph{IEEE Wireless
  Commun.}, vol.~31, no.~1, pp. 172--180, Dec. 2024.

\bibitem{li_coverage_2022}
J.~Li \emph{et~al.}, ``Coverage enhancement of {{5G}} commercial network based
  on reconfigurable intelligent surface,'' in \emph{{{Proc}}. {{IEEE VTC}}},
  Sep. 2022.

\bibitem{feng_deep_2020}
K.~Feng, Q.~Wang, X.~Li, and C.-K. Wen, ``Deep reinforcement learning based
  intelligent reflecting surface optimization for {{MISO}} communication
  systems,'' \emph{IEEE Wireless Commun. Lett.}, vol.~9, no.~5, pp. 745--749,
  May 2020.

\bibitem{feng_joint_2021}
K.~Feng, X.~Li, Y.~Han, and Y.~Chen, ``Joint beamforming optimization for
  reconfigurable intelligent surface-enabled {{MISO-OFDM}} systems,''
  \emph{China Commun.}, vol.~18, no.~3, pp. 63--79, Mar. 2021.

\bibitem{shaikh_performance_2022}
M.~H.~N. Shaikh, K.~Rabie, X.~Li, T.~Tsiftsis, and G.~Nauryzbayev, ``On the
  performance of dual {RIS-assisted} {V2I} communication under {Nakagami-m}
  fading,'' in \emph{{{Proc}}. {{IEEE VTC}}}, Sep. 2022.

\bibitem{ai_secure_2021}
Y.~Ai, F.~A.~P. {deFigueiredo}, L.~Kong, M.~Cheffena, S.~Chatzinotas, and
  B.~Ottersten, ``Secure vehicular communications through reconfigurable
  intelligent surfaces,'' \emph{IEEE Trans. Veh. Technol.}, vol.~70, no.~7, pp.
  7272--7276, Jul. 2021.

\bibitem{huang_transforming_2022}
Z.~Huang, B.~Zheng, and R.~Zhang, ``Transforming fading channel from fast to
  slow: {Intelligent} refracting surface aided high-mobility communication,''
  \emph{IEEE Trans. Wireless Commun.}, vol.~21, no.~7, pp. 4989--5003, Jul.
  2022.

\bibitem{9690475}
J.~Zhang \emph{et~al.}, ``{RIS}-aided next-generation high-speed train
  communications: {Challenges}, solutions, and future directions,'' \emph{IEEE
  Wireless Commun.}, vol.~28, no.~6, pp. 145--151, Dec. 2021.

\bibitem{dunna_scattermimo_2020}
M.~Dunna, C.~Zhang, D.~Sievenpiper, and D.~Bharadia, ``{ScatterMIMO}:
  {Enabling} virtual {MIMO} with smart surfaces,'' in \emph{{{Proc}}. {{ACM
  MobiCom}}}, Apr. 2020.

\bibitem{meng_rank_2023}
S.~Meng \emph{et~al.}, ``Rank optimization for {{MIMO}} systems with {{RIS}}:
  {{Simulation}} and measurement,'' \emph{IEEE Wireless Commun. Lett.},
  vol.~13, no.~2, pp. 437--441, Nov. 2024.

\bibitem{alexandropoulos_reconfigurable_2021}
G.~C. Alexandropoulos, N.~Shlezinger, and P.~{del Hougne}, ``Reconfigurable
  intelligent surfaces for rich scattering wireless communications: {{Recent}}
  experiments, challenges, and opportunities,'' \emph{IEEE Commun. Mag.},
  vol.~59, no.~6, pp. 28--34, Jun. 2021.

\bibitem{tang_wireless_2021}
W.~Tang \emph{et~al.}, ``Wireless communications with reconfigurable
  intelligent surface: {{Path}} loss modeling and experimental measurement,''
  \emph{IEEE Trans. Wireless Commun.}, vol.~20, no.~1, pp. 421--439, Jan. 2021.

\bibitem{tang_path_2022}
W.~\vspace{0mm}Tang \emph{et~al.}, ``Path loss modeling and measurements for
  reconfigurable intelligent surfaces in the millimeter-wave frequency band,''
  \emph{IEEE Trans. Commun.}, vol.~70, no.~9, pp. 6259--6276, Sep. 2022.

\bibitem{gao_propagation_2022}
B.~Gao \emph{et~al.}, ``Propagation characteristics of {{RIS-assisted}}
  wireless channels in corridors: {{Measurements}} and analysis,'' in
  \emph{{{Proc}}. {{IEEE ICCC}}}, Aug. 2022.

\bibitem{li_path_2022}
Y.~Li \emph{et~al.}, ``Path loss modeling for the {{RIS-assisted}} channel in a
  corridor scenario in {{mmWave}} bands,'' in \emph{{{Proc}}. {{IEEE
  GLOBECOM}}}, Dec. 2022.

\bibitem{zhang_channel_2023}
J.~Zhang \emph{et~al.}, ``Channel measurement, modeling, and simulation for
  {{6G}}: {{A}} survey and tutorial,'' 2023, arXiv:2305.16616. [Online].
  Available: http://arxiv.org/abs/2305.16616.

\bibitem{sun_3d_2021}
Y.~Sun, C.-X. Wang, J.~Huang, and J.~Wang, ``A {{3D}} non-stationary channel
  model for {{6G}} wireless systems employing intelligent reflecting surfaces
  with practical phase shifts,'' \emph{IEEE Trans. Cogn. Commun. Netw.},
  vol.~7, no.~2, pp. 496--510, Jun. 2021.

\bibitem{sang_multiscenario_2023}
J.~Sang \emph{et~al.}, ``Multi-scenario broadband channel measurement and
  modeling for sub-6 {{GHz RIS-assisted}} wireless communication systems,''
  \emph{IEEE Trans. Wireless Commun.}, early access, Nov. 2023.

\bibitem{hashemi_impulse_1993}
H.~Hashemi, ``Impulse response modeling of indoor radio propagation channels,''
  \emph{IEEE J. Sel. Areas Commun.}, vol.~11, no.~7, pp. 967--978, Sep. 1993.

\bibitem{cassioli_ultrawide_2002}
D.~Cassioli, M.~Win, and A.~Molisch, ``The ultra-wide bandwidth indoor channel:
  {{From}} statistical model to simulations,'' \emph{IEEE J. Sel. Areas
  Commun.}, vol.~20, no.~6, pp. 1247--1257, Aug. 2002.

\bibitem{lee_uwb_2010}
J.-Y. Lee, ``{{UWB}} channel modeling in roadway and indoor parking
  environments,'' \emph{IEEE Trans. Veh. Technol.}, vol.~59, no.~7, pp.
  3171--3180, Sep. 2010.

\bibitem{ichitsubo_multipath_2000}
S.~Ichitsubo, T.~Furuno, T.~Taga, and R.~Kawasaki, ``Multipath propagation
  model for line-of-sight street microcells in urban area,'' \emph{IEEE Trans.
  Veh. Technol.}, vol.~49, no.~2, pp. 422--427, Mar. 2000.

\bibitem{huang_channel_2021}
J.~Huang, C.-X. Wang, Y.~Yang, Y.~Liu, J.~Sun, and W.~Zhang, ``Channel
  measurements and modeling for 400\textendash 600-{{MHz}} bands in urban and
  suburban scenarios,'' \emph{IEEE Internet Things J.}, vol.~8, no.~7, pp.
  5531--5543, Apr. 2021.

\bibitem{7996408}
S.~Sun, H.~Yan, G.~R. MacCartney, and T.~S. Rappaport, ``Millimeter wave
  small-scale spatial statistics in an urban microcell scenario,'' in
  \emph{{{Proc}}. {{IEEE ICC}}}, Jul. 2017.

\bibitem{Molisch1999}
A.~F. Molisch and M.~Steinbauer, ``Condensed parameters for characterizing
  wideband mobile radio channels,'' \emph{Int. J. Wireless Inf. Netw.}, vol.~6,
  no.~3, pp. 133--154, Jul. 1999.

\bibitem{tang_estimation_2019}
P.~Tang, J.~Zhang, A.~F. Molisch, P.~J. Smith, M.~Shafi, and L.~Tian,
  ``Estimation of the {{K}}-factor for temporal fading from single-snapshot
  wideband measurements,'' \emph{IEEE Trans. Veh. Technol.}, vol.~68, no.~1,
  pp. 49--63, Jan. 2019.

\bibitem{saleh_statistical_1987}
A.~Saleh and R.~Valenzuela, ``A statistical model for indoor multipath
  propagation,'' \emph{IEEE J. Sel. Areas Commun.}, vol.~5, no.~2, pp.
  128--137, Feb. 1987.

\bibitem{karedal_measurementbased_2007}
J.~Karedal, S.~Wyne, P.~Almers, F.~Tufvesson, and A.~Molisch, ``A
  measurement-based statistical model for industrial ultra-wideband channels,''
  \emph{IEEE Trans. Wireless Commun.}, vol.~6, no.~8, pp. 3028--3037, Aug.
  2007.

\bibitem{ai_power_2015}
Y.~Ai, M.~Cheffena, and Q.~Li, ``Power delay profile analysis and modeling of
  industrial indoor channels,'' in \emph{{{Proc}}. {{IEEE EuCAP}}}, Apr. 2015.

\bibitem{meijerink_physical_2014}
A.~Meijerink and A.~F. Molisch, ``On the physical interpretation of the
  {{Saleh}}\textendash{{Valenzuela}} model and the definition of its power
  delay profiles,'' \emph{IEEE Trans. Antennas Propag.}, vol.~62, no.~9, pp.
  4780--4793, Sep. 2014.

\bibitem{czink_framework_2006}
N.~Czink, P.~Cera, J.~Salo, E.~Bonek, J.-P. Nuutinen, and J.~Ylitalo, ``A
  framework for automatic clustering of parametric {{MIMO}} channel data
  including path powers,'' in \emph{{{Proc}}. {{IEEE VTC}}}, Sep. 2006.

\bibitem{woon_identification_2006}
O.~H. Woon and S.~Krishnan, ``Identification of clusters in {{UWB}} channel
  modeling,'' in \emph{{{Proc}}. {{IEEE VTC}}}, Sep. 2006.

\bibitem{jiang_comparative_2020}
T.~Jiang, J.~Zhang, M.~Shafi, L.~Tian, and P.~Tang, ``The comparative study of
  {{S-V}} model between 3.5 and 28 {{GHz}} in indoor and outdoor scenarios,''
  \emph{IEEE Trans. Veh. Technol.}, vol.~69, no.~3, pp. 2351--2364, Mar. 2020.

\end{thebibliography}

\end{document}